\documentclass[aps,preprint,eqsecnum]{revtex4}
\usepackage{epsfig}
\begin{document}
\newcommand{\be}{\begin{eqnarray}}   
\newcommand{\ee}{\end{eqnarray}}
\newcommand{\ep}{\epsilon}
\newcommand{\si}{\sigma}

\tighten
\preprint{DO-TH 04/03}
\vspace*{1cm}
\title
{Accessing the Longitudinally Polarized Photon Content of the Proton}
\author{\bf A. Mukherjee}
\email{asmita@physik.uni-dortmund.de}
\author{C. Pisano}\email{pisano@harpo.physik.uni-dortmund.de}
\affiliation{ Institut f\"ur Physik, Universit\"at Dortmund, D 44221
Dortmund, Germany}
\date{\today\\[2cm]}

\begin{abstract}
We investigate the QED Compton process (QEDCS) in longitudinally polarized 
lepton-proton scattering
both in the elastic and inelastic channels and show that the cross section
can be expressed in terms of the polarized equivalent photon distribution of
the proton. We provide the necessary
kinematical constraints to extract the polarized photon content of the
proton using this process at  HERMES, COMPASS and eRHIC. We also discuss the
suppression of the major background process coming from virtual Compton
scattering. We point out that such an experiment can give valuable information on
$g_1(x_B, Q^2)$ in the small $x_B$, broad $Q^2$ region at the future
polarized collider eRHIC and especially in the lower $Q^2$, medium $x_B$
region in fixed target experiments.  
\end{abstract}
\maketitle

\section{Introduction}
QED Compton  process (QEDCS) in the scattering $l p \rightarrow l \gamma X$,
where $l$ is a lepton,  has a
distinctive experimental signature: both the outgoing lepton and photon
are detected at large polar angles and almost back to back in azimuth,
such that their transverse momenta almost balance each other, with little or
no  hadronic activity at the detector \cite{blu, ruju}. In fact, such a reaction in the unpolarized $ e-p $ scattering has long been suggested as an
excellent channel not only to determine the structure function $F_2(x_B,
Q^2)$ of the proton but also to extract the equivalent photon content of the
proton \cite{kessler,blu,ruju}. In a recent Monte Carlo analysis of the 
QED Compton
process performed by some members of the H1 collaboration at HERA \cite{lend},
it was found that, although the cross section in the elastic channel was
accurately described by the equivalent photon approximation (EPA), this was 
not the case in the
inelastic channel. In two previous papers \cite{pap1,pp2}  we have
suggested improved kinematical cuts for a more accurate extraction of
the unpolarized equivalent photon distribution of the proton which furthermore
suppress the major background process coming from virtual Compton scattering
(VCS). In this work we study the QED Compton process in the  polarized
scattering ${\vec l} {\vec p}\rightarrow l \gamma X $ (both elastic and
inelastic channels), where the initial lepton and proton are longitudinally polarized. We show that when the virtuality of the exchanged photon is
not too large, the cross section  can be approximated as a convolution of the 
longitudinally polarized equivalent photon
distribution of the  proton \cite{gpr1,gpr2} and the real photoproduction cross
section. We provide the necessary kinematical constraints to extract
the polarized photon content of the proton at HERMES, COMPASS and  eRHIC 
(the future polarized $ep$ collider planned at BNL). In addition, we
show that such an experiment can also access the polarized structure
function $g_1(x_B, Q^2)$ at HERMES in the low $Q^2$ region and at  eRHIC
over a wide range of the Bjorken scaling variable $x_B$ and $Q^2$.
$g_1(x_B, Q^2)$ and its first $x_B$  moment in the low $Q^2$ region have been 
studied in fully inclusive measurements at SLAC \cite{slac}, HERMES
\cite{her1,her2} and JLab \cite{clas1,clas2}. The most recent measurements by CLAS
\cite{clas3} are in the kinematical region $Q^2=0.15-1.64 ~~\mathrm{GeV}^2$.                   
The low $Q^2$ region is of particular interest because contributions due to
nonperturbative dynamics dominate here and thus the transition from soft to
hard physics can be studied. In fact the data in \cite{clas3} clearly
indicate a dominant contribution from the resonances and at higher $Q^2$
it is below the perturbative QCD  evolved scaling value of $g_1$. This in fact
illustrates the necessity of further investigation of $g_1 (x_B, Q^2)$ in the
transition region. In these fixed target experiments, low $Q^2$ is
associated with low values of $x_B$, thus the covered kinematical region is
smaller compared to the unpolarized data. Data on $g_1(x_B, Q^2)$ for small $x_B$ and
in the scaling region are missing due to the absence of polarized 
colliders so far (with the exception of RHIC, 
which has started operating in the polarized mode for $pp$ collisions 
only very recently). The small $x_B$ region is again interesting; 
it is the region
of high parton densities, and measurements in this region will provide
information about the effects of large $( \alpha_s ln^2{1\over x_B})^k$  
resummation and DGLAP evolution, and also about the 'soft' to 'hard' scale transition
\cite{badel,reya,bass}.
A better understanding of $g_1 (x_B, Q^2)$ in this region is necessary  
in order to determine its first moment experimentally. The kinematics of 
QED Compton events is 
different from the one of inclusive deep inelastic scattering 
due to the radiated photon in the final state and thus it provides a novel 
way to  access $g_1(x_B, Q^2)$ in a kinematical region not well covered by inclusive measurements (also for $F_2(x_B, Q^2)$ \cite{lend}).

The plan of the paper is as follows. In section II and III, we derive 
the analytic expressions of the cross section for the polarized QED 
Compton process in the elastic and inelastic channel, respectively. In
section IV, we discuss the background coming from virtual Compton
scattering (VCS) and also the interference between QEDCS and VCS. The
numerical results are presented in section V. Summary and conclusions are
given in section VI. The analytic expressions of the matrix elements are
given in Appendices A and B.        

\section{Elastic QED Compton scattering}
We consider QED Compton scattering in the elastic process:
\be
{\vec e}(l)+{\vec p}(P) \rightarrow e(l')+\gamma(k')+ p(P'),
\ee
where the incident electron and proton  are longitudinally polarized and 
the 4-momenta of the 
particles are given in brackets. Instead of the electron, one can also 
consider a  muon beam (COMPASS); the analytic expressions will be the same.  
We introduce the invariants 
\be
S=(P+l)^2, ~~~~~~~~~\hat s=(l+k)^2, ~~~~~~~~t=k^2.
\label{invar}
\ee
$k=P-P'$ is the 4-momentum of the virtual photon.
The photon in the final state is real, $k'^2=0$. We neglect the
electron mass everywhere except when it is necessary to avoid divergences in
the formulae and take the proton to be massive, $ P^2=P'^2=m^2 $. The
relevant Feynman diagrams for this process are shown in Fig. 1, with $X$
being a proton and $P_X=P'$. The
squared matrix element can be written as
\be
{\mid \Delta M^{QEDCS}_{\mathrm{el}} \mid }^2={1\over t^2} H^{A \mu \nu}_
{\mathrm{el}}(P,P') 
T^A_{\mu \nu}(l,k;l',k'),
\ee
$H^{A \mu \nu}_{\mathrm{el}}(P,P')$  and $T^A_{\mu \nu}(l,k;l',k')$ being the antisymmetric 
parts of the hadronic tensor and leptonic tensor respectively, which
contribute to the polarized cross section. As before \cite{pap1}, we use the
notation
\be
dPS_N(P;P_1,...,P_N)=(2 \pi)^4 \delta \bigg (P-\sum_{i=1}^N P_i \bigg ) 
\prod_{i=1}^N
{d^3P_i\over { (2 \pi)^3 2 P_i^0}}
\ee
for the Lorentz invariant $N$-particle phase-space element. The cross
section can be written as
\be
\Delta\sigma_{\mathrm{el}}(S)={1\over {2 (S-m^2)}} \int dPS_{2+1}(l+P;l',k',P')
{\mid \Delta M^{QEDCS}_{\mathrm{el}} \mid }^2~.
\label{sigmael}
\ee
Following the same approach as in \cite{kniehl,pap1} we can write this as
\be
\Delta\sigma_{\mathrm{el}}(S)={1\over {2 (S-m^2)}} \int {d\hat s\over 2 \pi}
\,dPS_2
(l+P;l'+k',P') {1\over t^2} H^{A \mu \nu}_{\mathrm{el}} (P,P') X^A_{\mu
\nu}(l,k)~.
\ee
$X^A_{\mu \nu}$ contains all the informations about the leptonic part of the
process and is defined as
\be
X^A_{\mu \nu}(l,k)=\int dPS_2(l+k;l',k') T^A_{\mu \nu}(l,k;l',k'),
\ee
$T^A_{\mu \nu}(l,k;l',k')$ is the antisymmetric part of the leptonic tensor,
\be
T_A^{\mu \nu}(l,k;l',k')=-{4  i e^4\over \hat s \hat u} \epsilon^{\mu
\nu \alpha \beta} k_\beta \Big [ (\hat s-t) l_\alpha+(\hat u-t) l'_\alpha
\Big ].
\label{lep}
\ee
Here $e^2=4 \pi \alpha$ and we have defined $\hat
t=(l-l')^2$ and $\hat u = (l-k')^2$.

For polarized scattering, $X^A_{\mu \nu}$ is antisymmetric in the indices
$\mu$, $\nu$  and can be expressed in terms of the  
Lorentz scalar $\tilde{X}^A_2$:
\be
X_A^{\mu \nu}=-{ i \over (\hat s -t)} \epsilon^{\mu \nu \alpha \beta}
k_\alpha l_\beta \tilde{X}^A_2(\hat s, t),
\label{xmunu}
\ee
with
\be
 \tilde{X}^A_2(\hat s, t)=- 2 X^A_{\mu \nu} P_A^{\mu \nu}.
\ee
$P_A^{\mu \nu}$ is the antisymmetric part of the photon polarization
density matrix \cite{flo}:
\be
P_A^{\mu \nu}~=~{1\over 2}  (\ep^\mu \ep^{\nu*}-\ep^\nu \ep^{\mu*})~=~
 {i\over 2 \sqrt {{| k^2 |}}} \ep^{\mu \nu \rho \si} k_{\rho} t_{\si} 
\ee
where $t_{\si}$ is the spin vector of the photon:
\be
t_{\si}= N_t \Big ( k_{\si}-{k^2\over l \cdot k} l_{\si}\Big );
~~~N_t={1\over \sqrt {{| k^2 | }}}.
\ee
We define the functions $X^A_2(\hat s, t ,\hat t) $ as
\be
\tilde{X}^A_2(\hat s,t)=2 \pi \int_{\hat t_{\mathrm{min}}}^
{\hat t_{\mathrm{max}}} d \hat t\,  {X}^A_2 (\hat s,t,\hat t).
\ee
The integration limits are given by Eq. (2.18) of \cite{pap1}.
$X^A_2(\hat s, t, \hat t)$ can be obtained using the leptonic tensor given
above:
\be
X^A_2(\hat s, t, \hat t)= {4 \alpha^2\over \hat s \hat u (\hat s-t)} \bigg [
{(\hat s -t)}^2+ {2 t \hat t (\hat u-t)\over \hat s-t} -  (\hat s +\hat t)^2
\bigg ].
\label{x2a}
\ee

The hadronic tensor for polarized scattering is expressed  in terms of the
proton form factors as \cite{ji,gpr1},
\be
H^A_{\mu \nu}=-i e^2 \ep^{\mu \nu \rho \si} m k_\rho \Big [ 2 G_E G_M S_\si
-{G_M (G_M-G_E)\over  1+\tau} {k\cdot S\over m^2} P_\si \Big ],
\ee
with $ \tau= {-t\over 4m^2}$ and $S_\si={1\over m} (P_\si -{m^2\over P \cdot l } l_\si)$ being the spin
vector of the proton which satisfies $S^2=-1 $ and $ P \cdot S=0$. $G_E$ and
$G_M$ are the proton electric and magnetic form factors and are empirically
parametrized as dipoles:
\be
G_E(t)= {1\over [1 - t/(0.71\,\mathrm{GeV}^2)]^{2}},~~~~G_M(t)=2.79~G_E(t).
\ee
The elastic cross section can then be written as
\be
\Delta \si_{\mathrm{el}} &=& {\alpha\over 8 \pi (S-m^2)^2} \int_{m_e^2}^{(
\sqrt{S}-m)^2} d\hat s \int_{t_{\mathrm{min}}}^{t_\mathrm{max}} {dt\over
t} \int_{\hat t_{\mathrm{min}}}^{\hat t_{\mathrm{max}}} d\hat t \int_{0}^{2 \pi} d \phi\,\, X^A_2(\hat s, t, \hat t) \nonumber\\&&~~\bigg [ \bigg ( 2 \,
{ S-m^2 \over \hat s-t }-1+{2 m^2\over t} {\hat s - t \over
 S-m^2} \bigg ) G_M^2(t)\nonumber\\&&~~-2\bigg ({ S-m^2 \over \hat s-t }-1+
{m^2\over t} {\hat s-t\over S-m^2} \bigg ) {G_M (G_M-G_E)\over 1+\tau}
\bigg ].
\label{elsigg}
\ee   
$\phi$ is the azimuthal angle of the outgoing $e-\gamma$ system in the
center of momentum frame.
The limits of integration are the same as in Eqs. (2.18) and (2.24) of
\cite{pap1}. These limits are modified due to the experimental cuts which we
impose numerically. In the EPA limit, we neglect $\mid t\mid $ vs.
$\hat s$ and $m^2$ vs. $S$ and get
\be
X_2^A(\hat s,t,\hat t)\approx X_2^A(\hat s,0,\hat t)~=~{4 \alpha^2\over \hat
s} \bigg ({\hat s\over \hat u}-{\hat u\over \hat s} \bigg )~=~ -{2 \hat s\over
\pi} {d \Delta \hat \si^{e \gamma \rightarrow e \gamma}\over d \hat t} ,
\ee 
where ${\Delta \hat \sigma \over d \hat t} $ is the  differential real 
photoproduction cross section 
and $ \tilde{X}_2^A(\hat s,0)=-4 \hat s \Delta \hat \si.$ The elastic cross section
then becomes
\be
\Delta\si_{\mathrm{el}} \approx \Delta\si_{\mathrm{el}}^{\mathrm{EPA}} = \int_{x_{min}}^{(1-{m\over \sqrt S})^2} dx
\int_{m_e^2-\hat s}^0 \,d\hat t\, \Delta \gamma_{\mathrm{el}}(x) {d\Delta \hat \si(xS,
\hat t) \over
d \hat t}
\ee
where $m_e$ is the mass of the electron and $\Delta \gamma_{\mathrm{el}}(x) 
$ is the elastic contribution to the polarized equivalent
photon distribution of the proton \cite{gpr1}
\be
\Delta \gamma_{\mathrm{el}}(x) = -{\alpha\over 2 \pi } \int_{t_{\mathrm{min}}}^{t_{\mathrm{max}}} {dt\over t} \Bigg [ \Big (2-x
+{2 m^2 x^2\over t} \Big ) G_M^2-2 \Big ( 1-x+ {m^2 x^2\over t}\Big ) {G_M
(G_M-G_E)\over 1+\tau} \Bigg ]\nonumber \\
\ee
with $x={\hat s\over S}$ and the limits of integration are given by Eq.
(2.30) of \cite{pap1}.
\section{Inelastic QED Compton Scattering}
We next consider the corresponding inelastic process
\be
\vec {e}(l)+\vec {p}(P) \rightarrow e(l')+\gamma(k')+X(P_{X}),
\label{eq2}
\ee
where $P_{X}=\sum_{X_i} P_{X_i}$ is the sum over all momenta of the produced
hadronic system. We take  the invariant mass of the produced hadronic system
to be $W$. The Bjorken variable $x_B$ is defined as 
\be
x_B= {Q^2\over 2 P\cdot (-k)} = {Q^2\over Q^2+W^2-m^2},
\ee
where $Q^2=-k^2=-t$. The cross section for inelastic scattering reads:
\be
\Delta \sigma_{\mathrm{inel}}(S)= {1\over 16 \pi^2 (S-m^2)^2} \int_{W^2_{\mathrm{min}}}^{W^2_{\mathrm{max}}} dW^2 \int_{m_e^2}^{(\sqrt{S}-W)^2} d \hat s 
\int_{Q^2_{\mathrm{min}}}^{Q^2_{\mathrm{max}}}
 {dQ^2\over Q^4} W_A^{\mu \nu} X^A_{\mu \nu} ,
\ee
where $X^A_{\mu \nu}$ is given by Eq. (\ref{xmunu}) and $W_A^{\mu \nu}$ is
the hadronic tensor:
\be
W_A^{\mu \nu}= i e^2 { m\over P \cdot k} 
\epsilon^{\mu \nu
\rho \sigma} k_\rho \bigg [ g_1(k^2, P \cdot k) S_\sigma +g_2 (k^2, P \cdot
k) \bigg (S_\sigma-{k \cdot S\over k \cdot P} P_{\sigma}  \bigg ) \bigg ] 
\ee
with $S_\sigma={1\over m } \Big ( P_\sigma - {m^2\over P \cdot l} l_\sigma \Big
) $ being the polarization of the proton, satisfying $S^2=-1$ and $P \cdot
S=0$.

The cross section can be written as
\be
\Delta \sigma_{\mathrm{inel}}(S) &=& {\alpha\over 4 \pi (S-m^2)^2} \int_{W^2_{\mathrm{min}}}^{W^2_{\mathrm{max}}} dW^2 \int_{m_e^2}^{(\sqrt{S}-m)^2} d \hat s \int_{Q^2_{\mathrm{min}}}^{Q^2_{\mathrm{max}}}
{dQ^2\over Q^2} {1\over (W^2+Q^2-m^2)} \nonumber \\ &&\,\times\,\,\bigg \{ \bigg [-2  {S-m^2\over
\hat s+Q^2}  +{W^2+Q^2-m^2\over Q^2}+{2 m^2 \over Q^2} \bigg ({\hat
s+Q^2\over S-m^2} \bigg ) \bigg ] g_1 (x_B, Q^2) \nonumber\\&&~~~~~+\,\,{4 m^2\over W^2+Q^2-m^2} g_2 (x_B, Q^2) \bigg \}
\tilde{X}_2^A(\hat s, Q^2),     
\ee
here $\tilde{X}_2^A (\hat s, Q^2) = 2 \pi\int_{\hat t_{\mathrm{min}}}^{\hat t_{\mathrm{max}}} d \hat t \, X_2^A(\hat s, Q^2, \hat t) $
with $X_2^A (\hat s, Q^2, \hat t)$ given by Eq. (\ref{x2a}).
The limits of the $Q^2, W^2$ and $\hat t$   integrations  are given in
Eqs. (3.11), (3.12) and (2.18) of \cite{pap1} respectively. In the limit of the EPA, as before, we
approximate $S-m^2 \approx S $ and $\hat s+Q^2 \approx \hat s $; 
the cross section then becomes
\be
\Delta \sigma_{\mathrm{inel}}(S) \approx \Delta \sigma_{\mathrm
{inel}}^{\mathrm{EPA}} =
\int_{x_{\mathrm{min}}}^{(1-m/\sqrt S)^2} dx \, \int_{m_e^2 -\hat s}^0   
d\hat t~ \Delta \gamma_{\mathrm{inel}}(x, x S) \,
\frac{ \Delta d\hat\sigma(x S, \hat t)}{d\hat t},
\label{epain}
\ee
where again $x={\hat s/S}$  and $\Delta \gamma_{\mathrm{inel}}(x, x S)$ is the
inelastic contribution to the polarized equivalent photon distribution of the
proton:
\be
\Delta \gamma_{\mathrm{inel}}(x, x S) = {\alpha\over 2 \pi} \int_x^1 {dy
\over y} \int_{Q^2_{\mathrm{min}}}^{Q^2_{\mathrm{max}}} {dQ^2\over Q^2} \bigg ( 2-y-{2m^2 x^2\over Q^2} \bigg ) 2 g_1 \bigg ({x \over y}, Q^2 \bigg ),
\label{gamin}
\ee
where we  have taken the scale to be $\hat s$. Here we
have neglected the contribution from $g_2(x_B, Q^2)$. Expressing $g_1(x_B, Q^2)$ in terms of the
polarized quark and antiquark  distributions, one can confirm that the above
expression reduces to that given in \cite{gpr1,gpr2}. However, in this case,
one chooses the minimal (but not compelling) boundary condition $\Delta
\gamma (x, Q_0^2) =0$ at a scale $Q_0^2=0.26 ~~\mathrm {GeV}^2$ . Eq. (\ref{gamin}) is free from
this particular boundary condition.  The limits of the $Q^2$
integration can be approximated similar to the unpolarized case (see Eq.
(3.16) of \cite{pap1}).     
\section{Background from Virtual Compton Scattering}
The cross section of the process in Eq. (\ref{eq2}) (also the elastic
channel) receives 
contribution from the
virtual Compton Scattering (VCS), when the photon is emitted from the proton
side (see Fig. 2) as well as the interference between the QED Compton
scattering (QEDCS) and VCS. The cross section for the elastic process is
given by
\be
\Delta \sigma_{\mathrm{el}}(S)={\alpha^3\over {8 \pi (S-m^2)^2}} 
\int_{m_e^2}^{(\sqrt{S}-m)^2} d\hat s \int_{{t}_{\mathrm{min}}}^{t_{\mathrm{max}}}
dt \int_{\hat t_{\mathrm{min}}}^{\hat t_{\mathrm{max}}} d\hat t \int_0^{2\pi} 
d\phi {1\over {(\hat s-t)}}
{{\mid {\Delta M_{\mathrm{el}}}\mid }^2},
\label{elsig}
\ee
where 
\be
{{\mid {\Delta M_{\mathrm{el}}}\mid }^2}= {{\mid
{\Delta M^{QEDCS}_{\mathrm{el}}}\mid
}^2}+{{\mid {\Delta M^{VCS}_{\mathrm{el}}}\mid }^2} - 2\, {\Re{\bf{\it e}}}\,
 {\Delta M^{QEDCS}_{\mathrm{el}}
\Delta M^{VCS *}_{\mathrm{el}}}
\ee
is the matrix element squared of the subprocess. 
The limits of integrations are the same as in Eq. (\ref{elsigg}). The 
interference term will have opposite sign if we consider a positron instead of
an electron. 
The cross section of the VCS process is expressed in terms of generalized
parton distributions and one needs a realistic model for a quantitative
estimate of this background \cite{gpd}. Here, in order to find the
constraints to suppress the VCS,
we make a simplified assumption: we take the proton to be a massive pointlike particle with an
effective $ \gamma^* p$ vertex, $ -i \gamma^\mu F_1(t) $. The explicit
expressions for the matrix elements are given in Appendix A.  

Particularly interesting for our purpose of extracting the polarized photon 
distribution of the proton is the inelastic channel. 
Here we use a unified parton model  (similar to our analysis in \cite{pp2}) 
to estimate the VCS and QEDCS rates. 
The cross section within the parton model is given by
\be
\frac{d \Delta \sigma_{\mathrm{inel}}}{d x_B \,d Q^2 \,d \hat s\, d \hat t \,d
\phi}=\sum_{q} \, \Delta q(x_B,Q^2)\, \frac{d \Delta \hat
\sigma^q}{d \hat s\, d Q^2 \,d \hat t\, d \phi},
\ee
where $\Delta q(x_B,Q^2)$  are the  polarized quark and antiquark 
distributions of the initial
proton, $q = u,\, d, \,s, \,\bar{u}, \,\bar{d}, \,\bar{s}$ and $d 
\Delta \hat \sigma^q$ is
the differential cross section of the subprocess
\be
{\vec e}(l)+{\vec q(p)} \rightarrow e(l')+\gamma(k')+q(p').
\label{eqsub}
\ee
Here ${\vec q}$ is a longitudinally polarized quark in a longitudinally
polarized proton and $q$ is a quark in the final state.
The integrated cross section becomes
\be
\Delta \sigma_{\mathrm{inel}}(S)&=& 
{\alpha^3\over {8 \pi (S-m^2)^2}} \sum_q
\int_{W^2_{\mathrm{min}}}^{W^2_{\mathrm{max}}}
dW^2 \int_{m_e^2}^{(\sqrt{S}-W)^2} d\hat s
\int_{Q^2_{\mathrm{min}}}^{Q^2_{\mathrm{max}}}
 {dQ^2 \over Q^2}
\int_{{\hat t}_{\mathrm{min}}}^{{\hat t}_{\mathrm{max}}}
d\hat t \int_0^{2\pi} d\phi {1\over {(\hat s+Q^2)}} 
\nonumber\\&&~~~~~~~~~~~~\times ~{{\mid {\Delta M_{\mathrm{inel}}}
\mid }^2} ~~\Delta q(x_B,Q^2),
\label{insig}
\ee
where
\be 
{{\mid {\Delta M_{\mathrm{inel}}}\mid }^2}= {{\mid
{\Delta M^{QEDCS}_{\mathrm{inel}}}\mid
}^2}
+{{\mid {\Delta M^{VCS}_{\mathrm{inel}}}\mid }^2} -2\, {\Re{\it e}} 
{\Delta M^{QEDCS}_{\mathrm{inel}}
\Delta M^{VCS *}_{\mathrm{inel}}}.
\ee 
The limits of integrations are the same as in Eq. (17) of \cite{pp2}.
The explicit expression of the matrix element is given in the appendix B.
It is useful to define the auxiliary invariants
$\hat S=(p'+k')^2$
and $\hat U= (p'-k)^2$, which can be written in terms of measurable 
quantities, 
\be
\hat S={\hat t (x_l-x_B)\over x_l}, ~~~~~~\hat U=\hat t- \hat S+Q^2,
\ee
with $x_l = {-\hat t\over 2 P \cdot (l-l')}$. In addition to the leptonic
variable $x_l$ we
define $x_\gamma={l\cdot k\over P\cdot l}$, which represents the fraction
of the longitudinal momentum of the proton carried by the virtual photon  
\cite{pap1}. In the limit of the EPA, both $x_l$ and $x_\gamma$  are the
same and become equal to $x$.
 
\section{Numerical Results}

In this section, we present our numerical results. The cuts used for HERMES,
COMPASS 
and eRHIC kinematics are given in Table I. The constraints on the energies and
polar angles of the detected particles reduce the background contributions
coming from the  radiative emissions (when the final state photon is emitted
along the incident or the final lepton line), because they prevent the
lepton propagators to become too small \cite{blu,thesis}. QED Compton 
events are singled out at HERA
by imposing a maximum limit on the acoplanarity angle $\phi_A$ which is
defined as $\phi_A=|\pi-\mid \phi_\gamma-\phi_e \mid |$, where $\phi_\gamma$
and $\phi_e$ are the azimuthal angles of the outgoing photon and electron,
respectively. We have observed \cite{pap1,pp2} that instead of this limit 
on $\phi_A$, the
constraint $\hat s > Q^2$, which is applicable experimentally,  is more efficient in extracting the equivalent
photon distribution from the 'exact' result. Here we use this constraint.

The unpolarized cross sections have been calculated using the formulae in
\cite{pap1} but for HERMES, COMPASS  and eRHIC kinematics, respectively.
In the numerical estimate of the unpolarized cross section, we have used
ALLM97 parametrization \cite{allm} of the structure function $F_2(x_B, Q^2)$ as
before, which is obtained by fitting DIS data of HERA and fixed target
experiments together with the total $p p$ and $\gamma p$ cross sections
measured and is expected to hold over the entire range of $x_B$ and $Q^2$.
We have taken $F_L(x_B, Q^2)$ to be zero, assuming  the Callan-Gross relation, similar to
\cite{thesis,lend}. In the polarized cross section, we have neglected the
contribution from $g_2(x_B, Q^2)$ and used the parametrization 
of \cite{bad} for
$g_1(x_B, Q^2)$. In this parametrization,  $g_1(x_B, Q^2)$ is described 
in the low-$Q^2$ region  by the GVMD
model together with the Drell-Hearn-Gerasimov-Hosoda-Yamamoto sum rule and the
asymptotic part of $g_1(x_B, Q^2)$ is expressed in terms of NLO GRSV00 \cite{grsv} parton
distributions (standard  scenario) in terms of a suitably defined scaling
variable ${\bar x}={Q^2+Q_0^2\over Q^2+Q_0^2+W^2-M^2}$ with $Q_0^2=1.2$
~~$\mathrm{GeV}^2$. The scale $Q^2$ is changed to $Q^2+Q_0^2$, so as to extrapolate to
low-$Q^2$ region. It is to be noted that for QED Compton scattering, the
effects of $F_L(x_B, Q^2)$ and $g_2 (x_B, Q^2)$ have to be taken into account 
in a more accurate
study as their effect may become non-negligible in the low-$Q^2$ region.
However, this is beyond the scope of the present work.

Before discussing the results for specific experiments, it is interesting
to investigate some general properties of the total cross section. Figs. 3
(a) and (b) show the total QEDCS cross section, polarized and unpolarized,
respectively, as a function of the incident lepton energy $E_l$. We have
imposed the constraints in the second column of Table I on the enegies and
angles of the outgoing particles, as well as those on $\hat s$. Both
polarized and unpolarized cross sections increase sharply with $E_l$, reach
a peak at around $E_l=20~\mathrm{GeV}$ and the start to decrease. The cross
section in the inelastic channel is also shown, which have  similar trends 
except that the peak in the polarized case is broader.  

\subsection{HERMES}
Figs. 4(a) and (b) show the total (elastic+inelastic) polarized and unpolarized QED Compton scattering
cross sections, respectively,  in bins of $x_\gamma$ for HERMES kinematics, subject to the
cuts of Table I. We have taken the incident electron energy $E_e=27.5$
$\mathrm{GeV}$. We also show the cross section calculated in the EPA. 
The same in the inelastic channel is also shown.
The cross section, integrated over $x_\gamma$, agrees with the EPA within $7.1\%$ (unpolarized) and
$4.8 \%$ (polarized). From the figures
it is also clear that the agreement in the inelastic channel is much better
than for HERA kinematics \cite{pap1,pp2} ($2.5 \%$ in the polarized case). 
This is because at HERMES $Q^2$
can never become too large (maximum $13.7$ ~$\mathrm{GeV}^2$), subject to our
kinematical cuts, which is expected in a fixed target experiment. 
The agreement is not so good without the constraint $\hat s >1$
~$\mathrm{GeV^2}$.  Fig. 4(c) shows the asymmetry, which is defined as
\be
A_{LL}={\sigma_{++}-\sigma_{+-}\over {\sigma_{++}+\sigma_{+-}}}
\ee
where $+$
and $-$ denote the helicities of the incoming electron and proton.  
They are calculated with the
same set of constraints. The asymmetry is quite sizable at HERMES and
increases in higher $x_\gamma$ bins. The asymmetry in the EPA is also shown.
It is interesting to note that the discrepancy in the cross sections with
the EPA approximated estimate, actually gets canceled in the asymmetry, as
a result it shows an excellent agreement with the EPA, except in the last
bin. We have also shown the expected statistical error in the bins, which
have been calculated using the formula, valid when the asymmetry is not too large:
\be
\delta A_{LL} \approx {1\over \mathcal {P}_e \mathcal{P}_p  \sqrt {\mathcal {L}
\sigma_{\mathrm{bin}}}};
\label{err}
\ee     
where $\mathcal {P}_e$ and $\mathcal {P}_p$ are the polarizations of the
incident lepton and proton, respectively, $\mathcal {L}$ is the integrated
luminosity and $\sigma_{\mathrm{bin}}$ is the unpolarized cross
section in the corresponding $x_\gamma$ bin. We have taken $\mathcal {P}_e=
\mathcal {P}_p=0.7$ and $\mathcal {L}={1 fb^{-1}}$ for HERMES. The expected
statistical error increases in higher $x_\gamma$ bins, because the number of
events become smaller. However the asymmetry seems to be measurable at HERMES.                

The background
from virtual Compton scattering is reduced at HERA by the experimental
condition of no observable hadronic activity at the detectors. Basically the
electron and photon are detected in the backward detectors and the hadronic
system in the forward detectors. In our previous work, we have observed that
for unpolarized scattering at HERA, such a constraint is insufficient to
remove the VCS contribution for higher $x_\gamma$. We have proposed a new
constraint $\hat S \ge \hat s  $, where $\hat S$ and $\hat s$ can be measured experimentally,  
to be imposed on the
cross section. Here, we investigate the effect of this constraint on the polarized cross section.  To estimate the inelastic contribution
coming from VCS, we use Eq. (\ref{insig}), together with an effective model
for the parton distribution of the proton. The effective parton distribution is of
the form 
\be
\Delta \tilde q(x_B, Q^2)= \Delta q({\bar x}, Q^2+Q_0^2),
\ee
$\Delta q (x_B, Q^2)$ being the NLO GRSV00 (standard scenario)  distribution 
function \cite{grsv}. In the relevant  kinematical 
region, $Q^2$ can be very small and may become close to zero, where the
parton picture is not applicable. The parameter $Q_0^2=2.3\,\,\mathrm{GeV}^2$ 
prevents the scale of the parton
distribution to become too small. ${\bar x}$ is a suitably defined scaling
variable, ${\bar x}= {x_B (Q^2+Q_0^2)\over Q^2+x_B Q_0^2}$. 
     
To estimate the unpolarized background
effect, we use the same expressions as in \cite{pp2} with an 
effective parton distribution given in Eq. (22) of
\cite{pp2}. Fig. 4(d) shows the polarized cross section in the inelastic
channel at HERMES, subject to the constraints of Table I, in bins of $\hat
s-\hat S$ calculated in the 'effective' parton model. The VCS and the
interference contributions are also shown. QEDCS cross sections using the
Badelek {\it et. al} parametrization of $g_1(x_B, Q^2)$ are also plotted. 
In fact, the
cross section in the 'effective' parton model lies close to this. Within the 
parton model,
VCS is suppressed when $\hat s < \hat S$, similar to the unpolarized case at
HERA \cite{pp2}. Unlike HERA, the interference between QEDCS and VCS is not
negligible at HERMES, although smaller than QEDCS in the relevant region.
Since the interference term changes  sign when a  positron beam is used instead
of the electron beam, a combination of electron and positron scattering data can
eliminate this contribution. In order to estimate the VCS in the elastic
channel, one needs a suitable model for the polarized generalized parton
distributions. However, in the simplified approximation of a pointlike
proton  with an
effective vertex as described in section IV, the elastic VCS as well as the 
interference contribution is
much suppressed at HERMES. Similar observations hold for unpolarized
scattering.

Fig. 5 shows the asymmetries in the inelastic channel in bins of $x_B$.
In addition to the cuts mentioned above and shown in table I, we have also chosen $\hat S -\hat s
> 2~~\mathrm{GeV}^2$ to suppress the background. The asymmetry is small but sizable and could
be a tool to access $g_1(x_B, Q^2)$ at HERMES. In fact, QED 
Compton events can be
observed at HERMES in the kinematical region $x_B=0.02-0.7$ and
$Q^2=0.007-7~ \mathrm{GeV}^2$ (small $Q^2$, medium $x_B$).  
However, from the figure 
it is seen that the asymmetry
is very small for $x_B$ below $0.1$. We have also shown the expected
statistical error in each bin. The average $Q^2$ value in $\mathrm{GeV}^2$ for the
polarized cross section for each
bin is shown, which has been calculated using the formula
\be
\langle Q^2\rangle={\int_{\mathrm{bin}} Q^2\, d \Delta \sigma \over \int_{\mathrm{bin}} d \Delta \sigma}
\ee       

\subsection{COMPASS}

Figs. 6 (a) and (b) show the cross sections in bins of $x_\gamma$ for the 
polarized and 
unpolarized QEDCS for the kinematics of COMPASS. We take the energy of the
incident muon beam to be  $160 \,\,\mathrm{GeV}$, the target is a proton. 
The final muon and the photon are detected
in the polar angle region $ 0.04 < \theta_\mu,
\theta_\gamma < 0.18$. The cross sections in bins, subject to the kinematical 
constraints shown in Table I, are much smaller than at HERMES, because 
they start to decrease with the increase of the incident lepton energy 
$E_l$ as $E_l$ becomes greater than about $20$ $\mathrm{GeV}$, as observed
in Fig. 3. 
As before, the cuts remove the initial and final state radiative events. The
$x_\gamma$ integrated cross section agrees with the EPA within $14.2 \%$
(unpolarized) and $15.5 \%$ (polarized). The agreement thus is not as good
as at HERMES. From the figures it is seen that the coss section in the EPA 
actually lies below
the 'exact' one, both for polarized and unpolarized cases. This discrepancy
is due to the fact that the EPA is expected to be a good approximation when
the virtuality of the exchanged photon is small. At COMPASS, with our
kinematical cuts, $Q^2$ can not reach a value below $0.07 \,\,\mathrm{GeV}^2$ and can
be as large as $144\,\,\mathrm{GeV}^2$, whereas for HERMES smaller values of $Q^2$ are
accessible (see the previous subsection). Fig. 6(c) shows the
asymmetry in bins of $x_\gamma$, the asymmetry in the inelastic channel is
also shown. The asymmetry is of the same order of magnitude as in HERMES and
is in good agreement with the EPA. We have also shown the expected
statistical error in each bin, calculated using  Eq. (\ref{err}). We have
taken  $\mathcal {P}_e=
\mathcal {P}_p=0.7$ and $\mathcal {L}={1fb^{-1}}$ for COMPASS. The statistical
error is large in higher $x_\gamma$ bins. Fig. 6(d) shows the polarized
QEDCS, VCS and interference contributions (inelastic) calculated in the 'effective'
parton model, in bins of $\hat s-\hat S$. As in HERMES, VCS is suppressed for
$\hat s < \hat S$. The interference term is not suppressed but using $\mu^+$
and $\mu^-$ beams this can be eliminated. We have also plotted the QEDCS
cross section using Badelek {\it et} {\it al.} parametrization of $g_1(x_B,Q^2)$.
The VCS and the interference contributions (elastic) are much suppressed in the
pointlike approximation of the proton with the effective vertex.
Fig. 7 shows the asymmetry at COMPASS in the inelastic channel plotted in
bins of $x_B$ with the same set of constraints and the additional cut $\hat S - \hat s > 2 \,\,\mathrm{GeV}^2$. The asymmetry is sizable and
can give access to $g_1(x_B, Q^2)$, the kinematically allowed range is $0.07 < x_B$ . 
We have also shown the expected statistical errors in the
bins and the average $Q^2$ in each bin. Comparing Fig. 5
and 7 one can see that there is no overlap in the kinematical region covered 
at HERMES and COMPASS. Higher values of $Q^2$ are probed at COMPASS in the 
same $x_B$ range as compared to HERMES.             
   
\subsection{eRHIC}

The cross sections for eRHIC kinematics, both polarized and unpolarized, are 
shown in Fig. 8(a) and (b), respectively,
in bins of $x_\gamma$. We have taken the incident electron energy $E_e=10 $
GeV and the  incident proton energy $E_p= 250$ GeV. The cross section in the
EPA is also shown. The kinematic constraints are given in Table I.
The polar angle acceptance of the detectors at eRHIC is not known. We have
taken the range of $\theta_e, \theta_\gamma $ to be the same as at HERA. We
have checked that the constraints on the energies and the polar angles of
the outgoing electron and photon are sufficient to prevent the electron
propagators to become too small and thus reduce the radiative contributions.
The unpolarized total (elastic+inelastic) cross section, 
integrated over $x_\gamma$
agrees with the EPA within $1.6 \%$. The agreement in the inelastic channel
is about $6.3 \%$. The polarized total cross section agrees with the 
EPA within $9.8 \%$. The
EPA in this case lies below the 'exact' one in all the bins. The agreement in
the inelastic channel is about $19.6\%$. More restrictive constraints
instead of $\hat s > Q^2$, like $\hat s > 10~ Q^2$, makes the agreement
better, about $1.2 \%$ in the polarized case and $1.9 \%$ in the unpolarized
case. Fig. 8(c) shows the asymmetry for eRHIC, in bins of $x_\gamma$. The
discrepancy in the cross section cancels in the asymmetry, as a result good
agreement with the EPA is observed in all  bins except the last one at
higher $x_\gamma$. The asymmetry in the inelastic channel is also shown. We
have plotted the expected statistical error in the bins using Eq.
(\ref{err}). For eRHIC, we have taken  $\mathcal {P}_e=
\mathcal {P}_p=0.7$ and $\mathcal {L}={1fb^{-1}}$. The expected statistical error  
increases in higher $x_\gamma$ bins. The asymmetry is very small for small
$x_\gamma$ but becomes sizable as $x_\gamma$ increases. Fig. 8(d) shows the
polarized cross section in the inelastic channel, in bins of $\hat s-\hat S$,
in the 'effective' parton model for eRHIC. The VCS is suppressed in all
bins, especially for $\hat s <\hat S$. The interference contribution is
negligible, similar to HERA. The 'effective' parton model QEDCS cross
section is also compared with the more exact one, using Badelek {\it et. al.}
parametrization for $g_1(x_B, Q^2)$. 
Similar effects are observed in the unpolarized
case. In the pointlike approximation of the proton with the effective vertex, 
as before, the elastic VCS as well as
the interference contributions are very much suppressed. Fig. 9 shows the 
asymmetry in bins of $x_B$ in the inelastic channel, which may be relevant for
the determination of $g_1(x_B, Q^2)$ using QEDCS at eRHIC. The asymmetry is small but
sizable, however the error bars are large and therefore good statistics is
needed. $x_B$ can be as low as $0.002$. A wide range of $Q^2$ can be
accessed at eRHIC starting from $0.008$ to $2000$ $\mathrm{GeV}^2$; 
the average $Q^2$ value in the bins ranges from $2.4$ to $315$ $\mathrm{GeV}^2$. Fig. 10
shows the total  asymmetry in $Q^2$ bins for eRHIC. The asymmetry in this case
is bigger in each bin and the error bars are smaller  than the $x_B$ bins except the last bin
for high $Q^2$ where the number of events are smaller.     
\section{Summary and Conclusions}

To summarize, in this paper we have analyzed the QED Compton process in
polarized $lp$ scattering, both in the elastic and inelastic channel. This
process has a distinctive experimental signature and we showed that the 
cross section can
be expressed in terms of the equivalent photon distribution of the polarized 
proton,
convoluted  with the real photoproduction cross section. The EPA is a useful
tool to estimate high energy scattering cross sections; however the accuracy
of this approximation and the kinematical region of its validity have to be
checked by experiment. In this work we provided the necessary kinematical
constraints for the extraction of the polarized photon content of 
the proton by measuring the 
QED Compton process at HERMES, COMPASS and
eRHIC. We showed that the
cross section and, in particular, the asymmetries are quite accurately 
described
by the EPA. We also discussed the possibility of suppressing  the major
background process, namely the virtual Compton scattering. 
We pointed out that such an
experiment can give access to the spin structure function $g_1(x_B, Q^2)$ in
the region of low $Q^2$ and medium $x_B$ in fixed target experiments and
over a broad range of $x_B,~ Q^2$ at the future polarized $ep$ collider,
eRHIC. Because of the different kinematics compared to the fully inclusive
processes, the QED Compton process can provide information on $g_1(x_B, Q^2)$ 
in a range
not well-covered by inclusive measurements and thus is a valuable tool to
have a complete understanding of the spin structure function.         
\section{acknowledgements}
We warmly acknowledge E. Reya and M. Gl\"uck for initiating  this study, as
well as for  many fruitful discussions. We also thank W.
Vogelsang for
helpful discussions and suggestions. This work has been supported in part by
the 'Bundesministerium f\"ur Bildung und Forschung', Berlin/Bonn.

\appendix
\section{Matrix Element for the Elastic Background Virtual Compton
Scattering Process}

The explicit expressions of the matrix elements in section IV are given below:
\be
{{\mid {\Delta M^{QEDCS}_{\mathrm{el}}}\mid }^2} &=&   \frac{4}{t \,\hat{s}\, \hat u}\, \bigg [ -A + \frac{2\,m^2}{t\, S'} \,B \bigg ]\, F_1^2(t), 
\ee
\be
{{\mid {\Delta M^{VCS}_{\mathrm{el}}}\mid }^2} &=& -\frac{4}{\hat t \,U'\, \hat{S}'}\, \bigg [ A + \frac{2\,m^2}{\hat {S}'\, S' U' } \,C \bigg ]\, F_1^2(\hat t), 
\ee
with
\be
A  =  2 t^2 + (\hat s - 2 S' - \hat u) (\hat s + \hat u) - 2 t (\hat s - 
2 S' - U') - 2 \hat u U',
\ee
\be
B =  - 2 t^3  + \hat{s}^3 - \hat s \hat {u}^2 + 2 t^2 (2 \hat s + \hat u) - t 
(3 \hat {s}^2 + \hat{u}^2),
\ee
\be
C  & = & (\hat s + \hat u)^2 \,[-2 S'^2 + \hat s \hat u - 2 S' \hat u - \hat{u}^2 
+ 2 m^2 ( \hat {s} + \hat {u}- t) - 
     t (S' + \hat{u})]  \nonumber \\ & & \,\,- (\hat{s} + \hat{u}) \,[2 t^2 - 3 t \hat{s} + \hat{s}^2 + \hat{s} (S' - 2 \hat u) + 
     3 \hat u (S' + \hat u)]\, U'\nonumber \\ &&\,\,\,\,- \,[2 t^2 + s^2 - \hat s \hat u + 2 \hat{u}^2 - t (3 \hat s + \hat u)] \,U'^2.
\ee
\be
 2\, {\Re{\bf{\it e}}}\,
 {\Delta M^{QEDCS}_{\mathrm{el}}\Delta M^{VCS *}_{\mathrm{el}}} &=& -\frac{4}{ t\, \hat s \, \hat u\, \hat t\, U'\, \hat{S}'}   \,\bigg [ {D + \frac{2 m^2 E}{S'}} \bigg ]\,F_1(\hat t) \,F_1(t),  
\label{qel}
\ee
where $D$ and $E$ read:
\be
D & = & [ 2 t^2 + (\hat s - 2 S' - \hat u) (\hat s + \hat u) - 2 t (\hat s - 2 S' - U') -2 \hat u U' ]
 \nonumber \\
 & & \,\,\times  \{ (\hat s + \hat u) [t \hat u + S' (\hat s + \hat u)] + 
 [t  (\hat s - \hat u) + \hat s (\hat s + \hat u)] U'\},
\ee
\be
E & = & [\hat s ( \hat s - t)^2 (\hat s - 2 t) + ( 2 t ^3  - t^2  \hat s - \hat{s}^3) \hat u + (-2 t^2  - 3 t \hat{s} + \hat{s}^2) \hat{u}^2 + ( t +  3 \hat s) 
\hat{u}^3]\, U'\nonumber \\ 
 & & - (\hat s + \hat u) \{-2 t^3 \hat u - (\hat s + \hat u) 
[\hat{s}^2 (S' - 2 \hat{u}) + S' \hat{u}^2 + 2 \hat{s} \hat{u} (S' + \hat u)] \nonumber \\  &&
 \,\,- t [-7 \hat{s} S' \hat{u} + \hat{u}^2 (-2 S' + \hat u)  + 
        \hat{s}^2 (-3 S' + 5 \hat u)] + t^2 [2 \hat{u} ( \hat u -S') + \hat s 
( 5 \hat u - 2 S')] \}.\nonumber \\
\ee

We have introduced the invariants $ U=(P-k')^2, ~\hat u=(l-k')^2
$  and $\hat S=-(\hat s+\hat u+U'-m^2)$
and used the notations $S'=S-m^2, ~U'=U-m^2,~\hat S'=\hat S-m^2$  
for compactness.

\section{Matrix Element for the Inelastic Virtual Compton Scattering 
Background process}

For the corresponding inelastic channel in section IV the explicit matrix elements 
are given by:
\be
{{\mid
{\Delta M^{QEDCS}_{\mathrm{inel}}}\mid}^2}  &=&  4 \,e_q^2 \,\frac{F}{Q^2\,\hat s \,\hat u}.
\ee
\be
{{\mid {\Delta M^{VCS}_{\mathrm{inel}}}\mid }^2} & = & - 4 \,e_q^4 \,\frac{F}{\hat t\,\hat S \,\hat U},
\ee
with $\hat S = - (\hat s + \hat u + x_B U')$, $\hat U = x_B U'$ and
\be 
F & = & \hat{s}^2 - \hat{u}^2 + 2 Q^4 + 2 Q^2 \hat{s} - 
  2 x_B [\hat{s} S' + \hat{u} (S' + U') + Q^2 (2 S' + U')].
\ee
Here $e_q$ is the charge of the parton in units of the proton charge. Finally
\be
2\, {\Re{\it e}} 
{\Delta M^{QEDCS}_{\mathrm{inel}}
\Delta M^{VCS *}_{\mathrm{inel}}} &=&  \,-4\, e_q^3\, \frac{G\, H}{Q^2\,\hat s\, \hat u 
\,\hat t \,\hat S\, \hat U},
\ee
with
\be
G & = & 2 Q^4 + \hat{s}^2 - 2 \hat s S' x_B - \hat u [\hat u + 2 (S' + U') 
x_B] +    2 Q^2 [\hat s - (2 S' + U') x_B]
\ee
\be
H = Q^2 [\hat s (\hat u - \hat U) + \hat u (\hat u + \hat U)].   
- x_B\,(\hat s + \hat u)[S' \hat u + \hat s (S' + U')]
\ee 

 
\newpage
\vspace*{.2cm}
\begin{center}
\epsfig{figure= 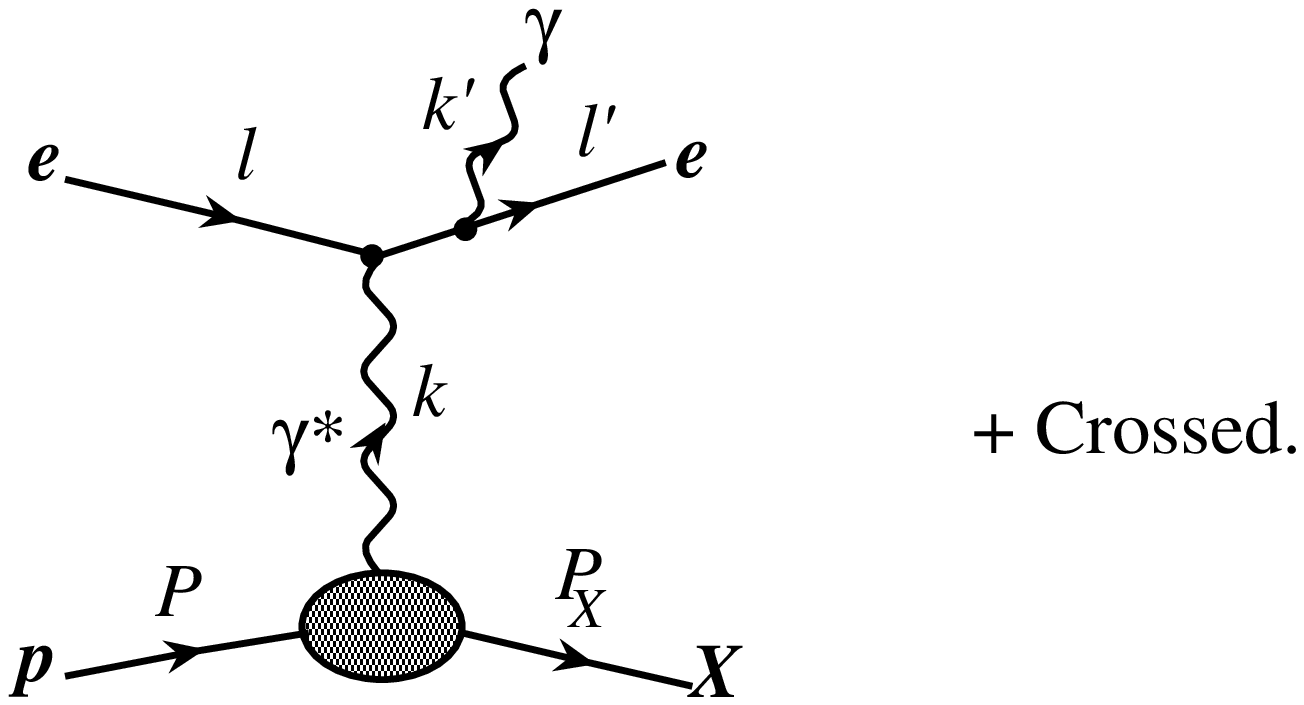, width=16cm, height= 6cm}\\
\end{center}
\begin{center}
\parbox{10.0cm}
{{\footnotesize
 Fig. 1:  Feynman diagrams for the QED Compton process (QEDCS). $X \equiv p$
(and $P_X \equiv P'$) 
corresponds to elastic scattering.}}
\end{center}
\vspace{1cm}
\begin{center}
\epsfig{figure= 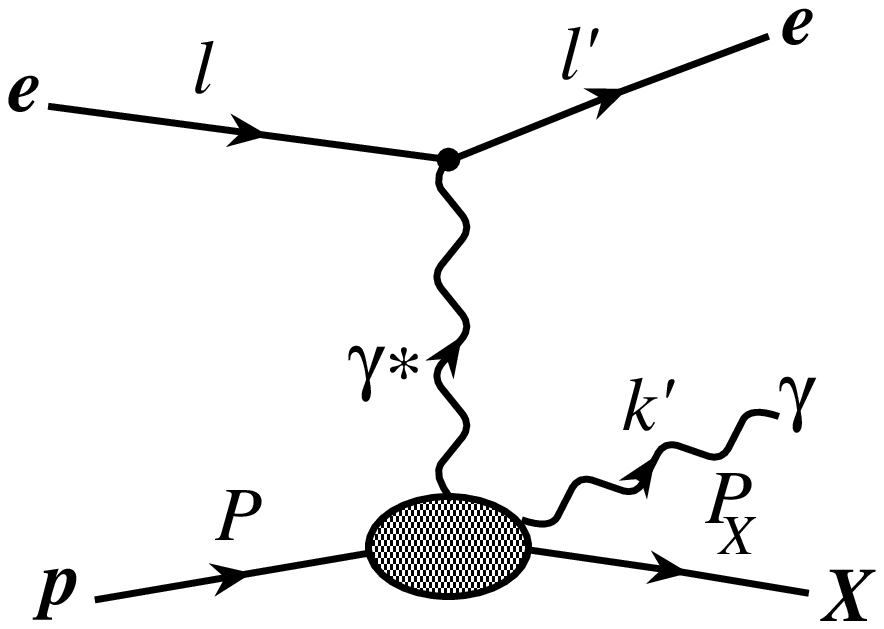, width=16cm, height= 6cm}\\
\end{center}
\begin{center}
\parbox{10.0cm}
{{\footnotesize
 Fig. 2:  As in Fig. 1 but for the virtual Compton scattering (VCS)
background process. }}
\end{center}
\newpage
\begin{center}
\parbox{8cm}{\epsfig{figure=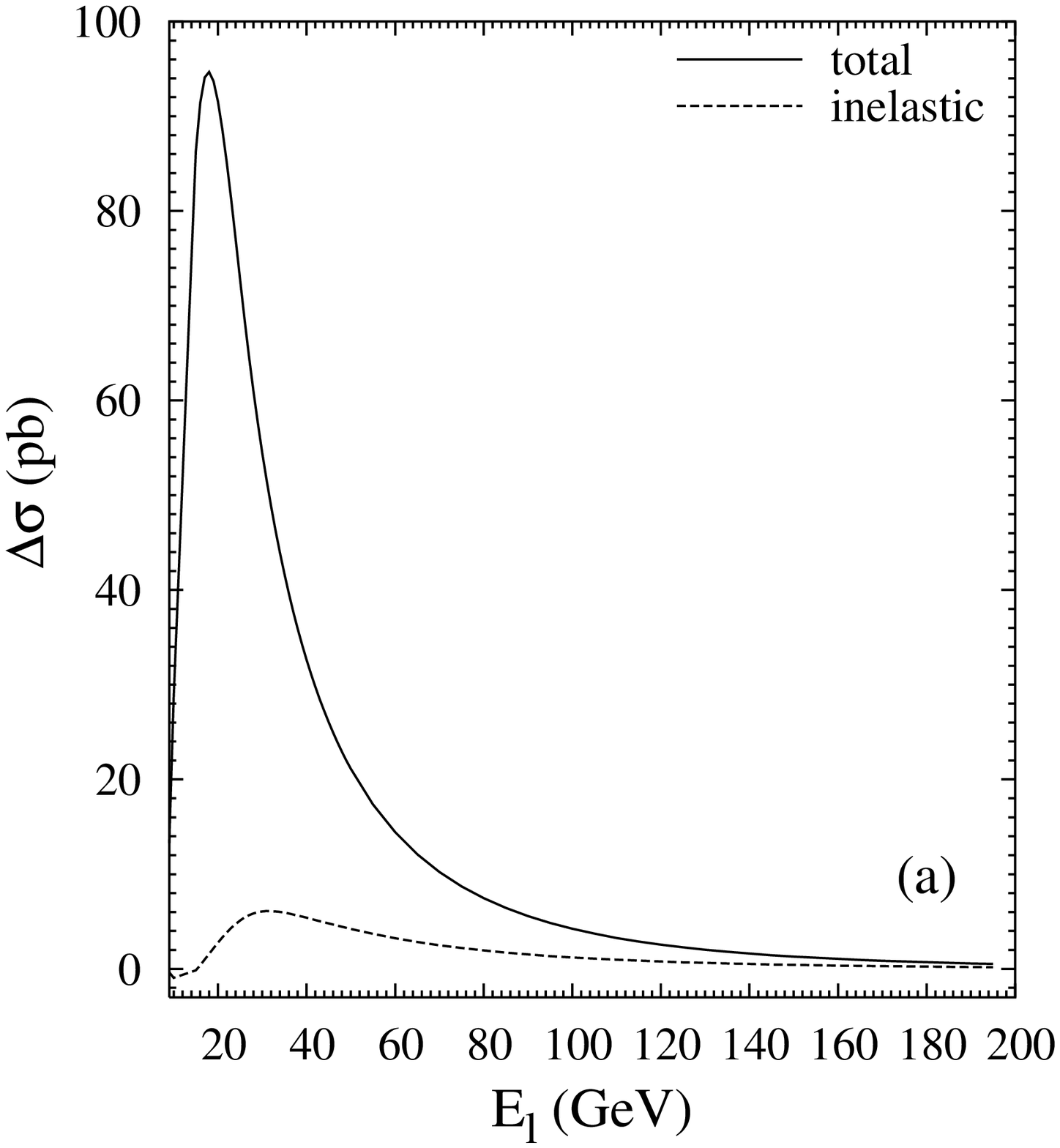,width=8.5 cm,height=7.5 cm}}\
\
\parbox{8cm}{\epsfig{figure=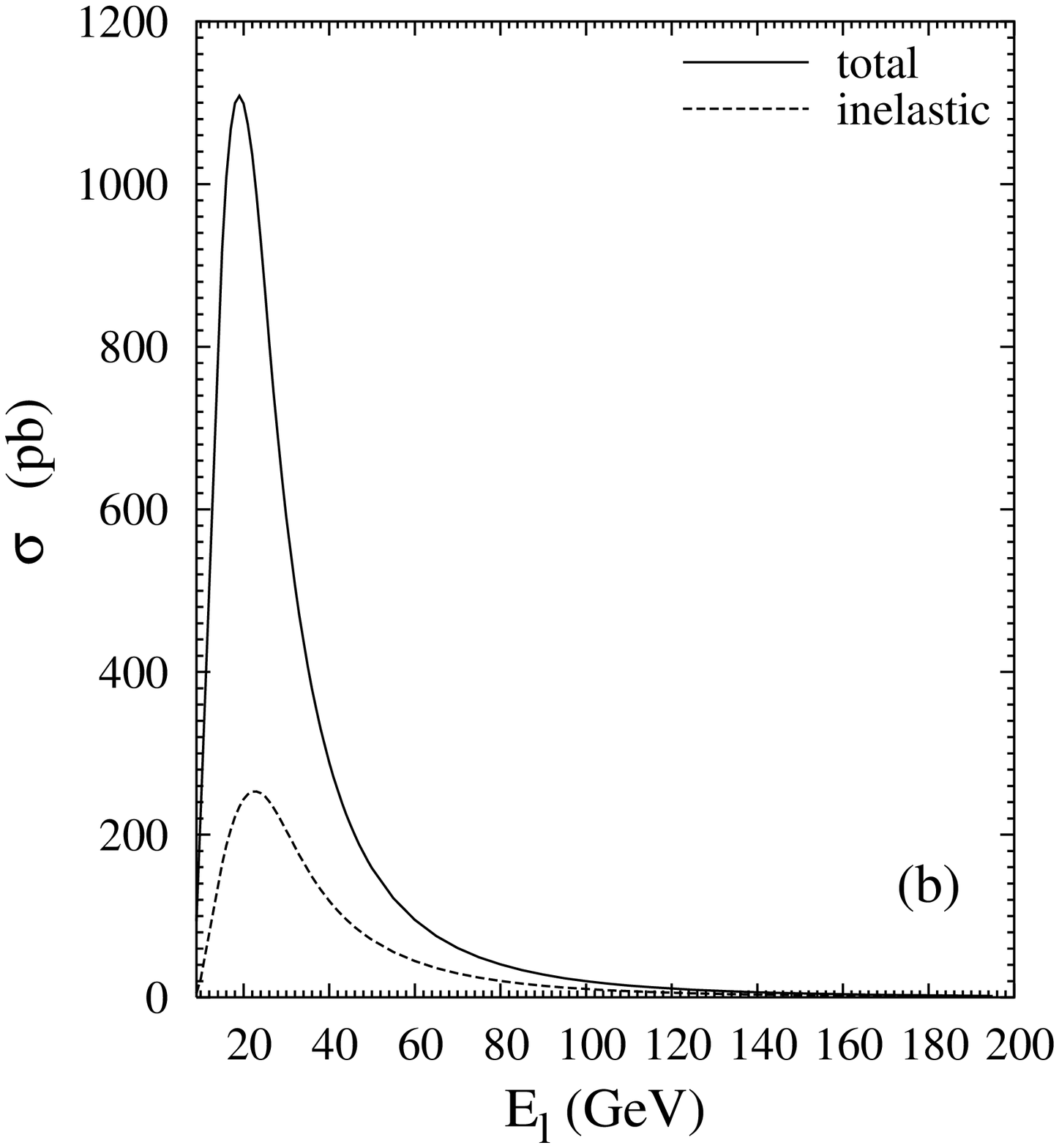,width=8.5 cm,height=7.5 cm}}\
\
\end{center}   
\vspace{0.2cm}
\begin{center}
\parbox{14.0cm}
{{\footnotesize
Fig. 3: QEDCS cross section vs. energy of the incident lepton; (a)
polarized, (b) unpolarized. The continuous line is the total cross section and 
the dashed line is the cross section in the inelastic channel. The cuts imposed
are given in the central column  of table I. 
We have used the ALLM paramtrization of
$F_2$ \cite{allm} and the Badelek {\it et. al} parametrization of $g_1$
\cite{bad}.}}
\end{center}

\newpage      
\begin{center}
\parbox{8cm}{\epsfig{figure=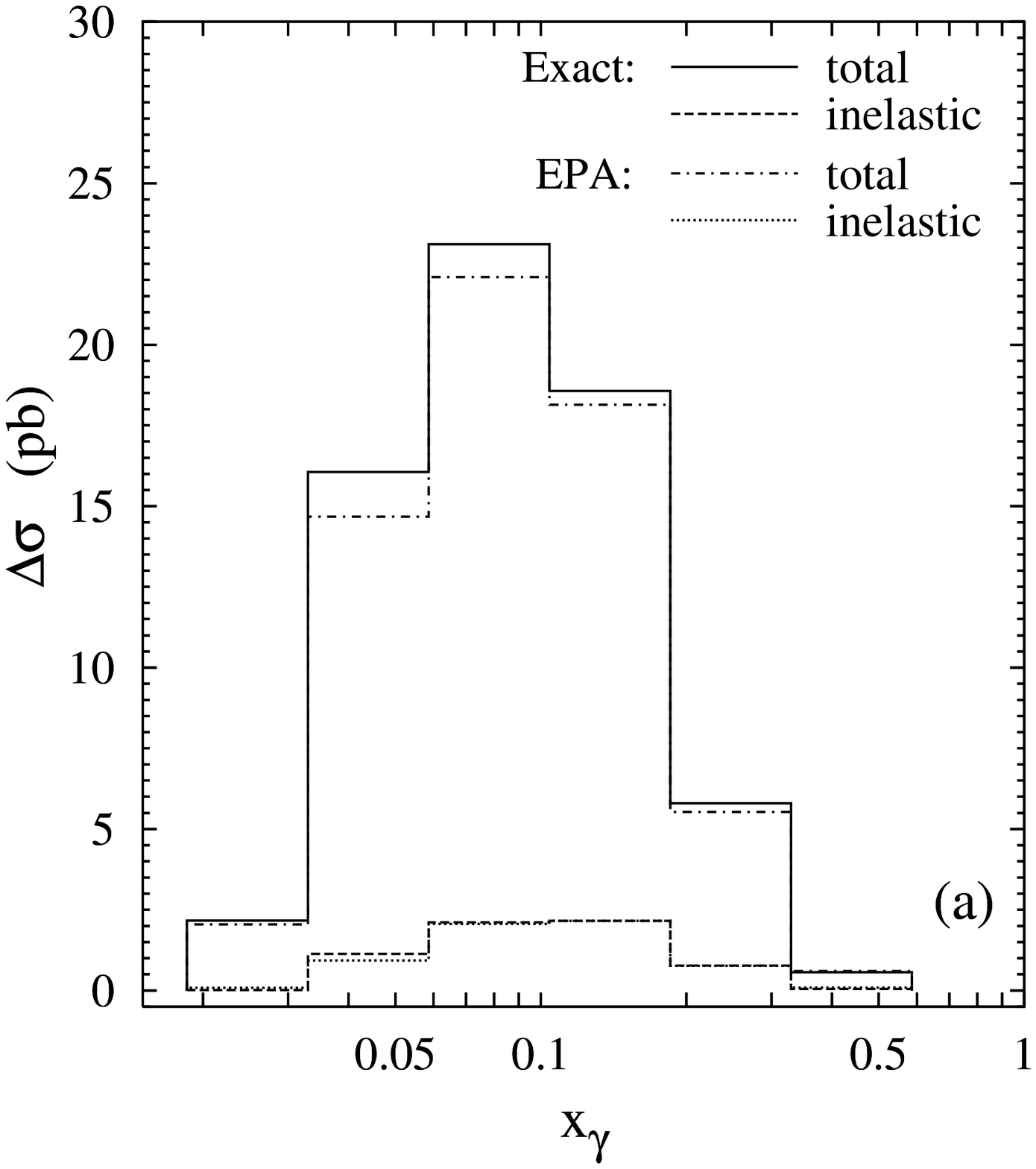,width=8.5 cm,height=7.5 cm}}\
\
\parbox{8cm}{\epsfig{figure=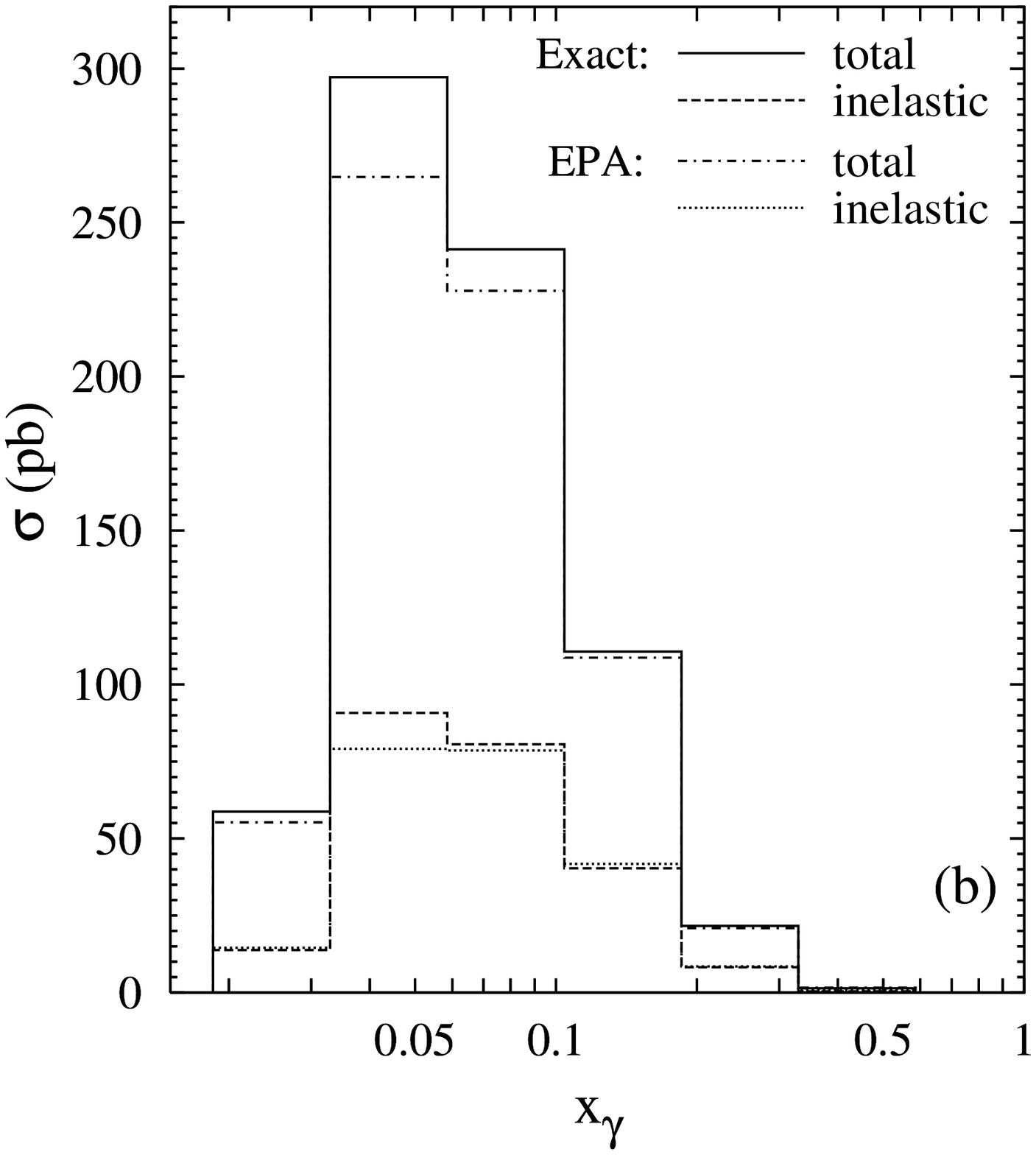,width=8.5 cm,height=7.5 cm}}\
\
\end{center}   
\vspace{0.2cm}

\begin{center}
\parbox{8cm}{\epsfig{figure=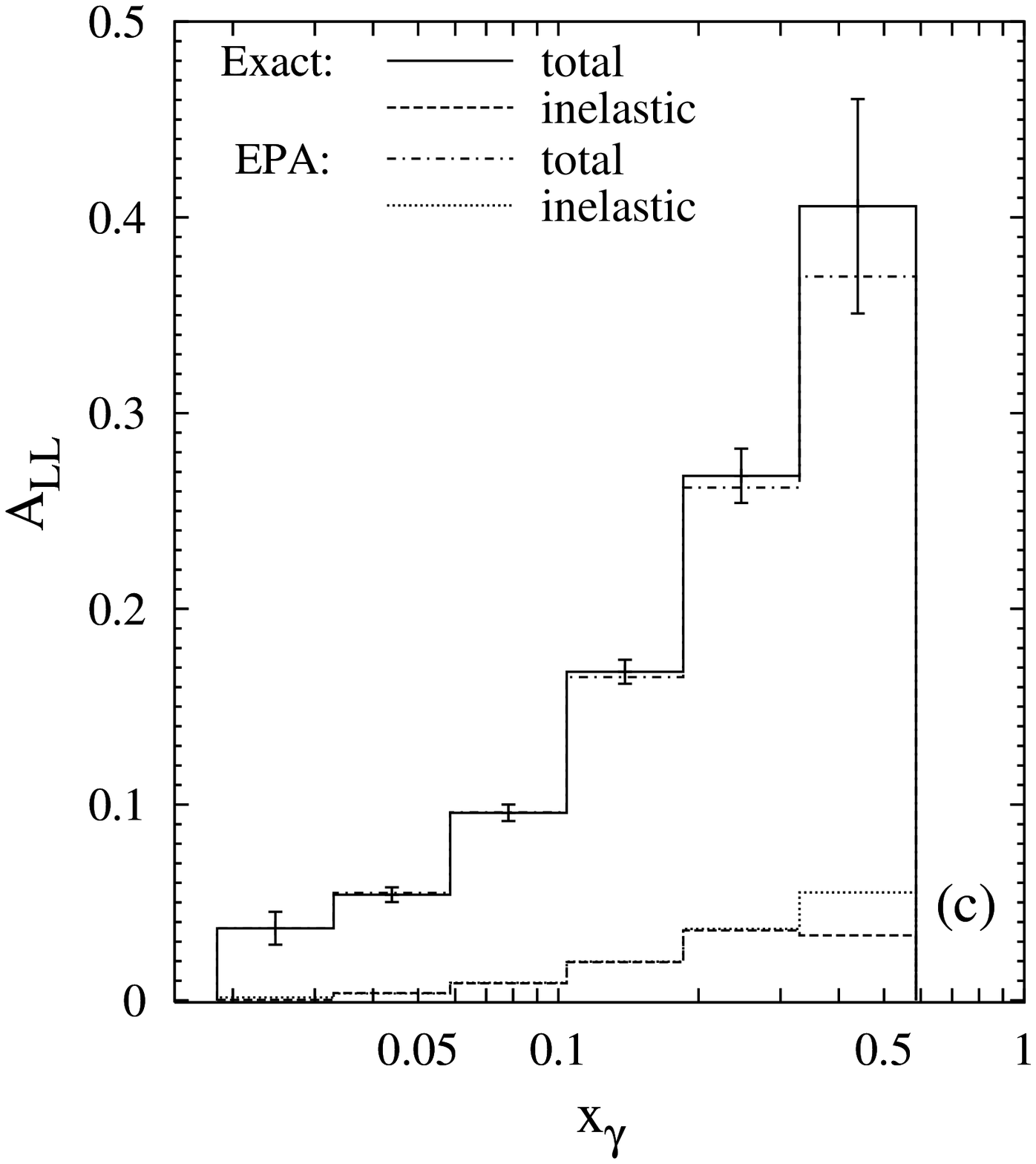,width=8.5 cm,height=7.5 cm}}\
\
\parbox{8cm}{\epsfig{figure=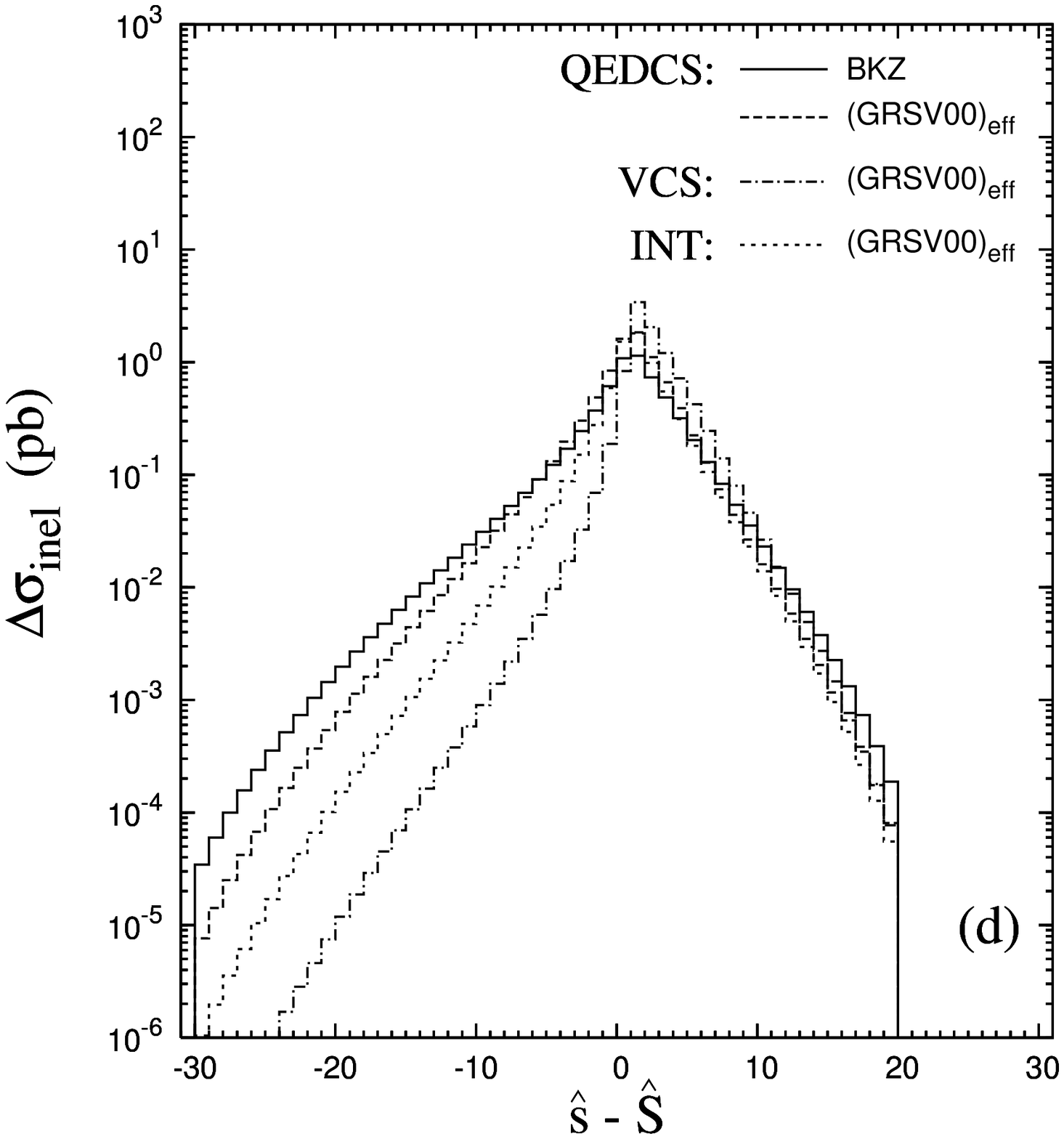,width=8.5 cm,height=7.5 cm}}\
\
\end{center}   
\vspace{0.2cm}
\begin{center}
\parbox{14.0cm}
{{\footnotesize
Fig. 4:  Cross section for QED Compton scattering (QEDCS) at
HERMES in bins of $x_\gamma$ (a) polarized, (b) unpolarized, 
(c) the asymmetry; for the polarized cross section Badelek {\it et al.} 
\cite{bad} parametrization of $g_1$ (BKZ) and for the unpolarized cross section ALLM
parametrization of $F_2$ have been used; (d) polarized inelastic cross section 
for QEDCS (long dashed), VCS
(dashed-dotted) and the interference (dashed)  at 
HERMES in the effective parton model.
The bins are in $\hat s- \hat S$, expressed in $\mathrm{GeV^2}$. The
continuous  line is the QEDCS cross section using 
the BKZ parametrization of $g_1(x_B, Q^2)$. The constraints
imposed are given in Table I.}} 
\end{center}
\newpage
\vspace*{4cm} 
\begin{center}
\epsfig{figure= 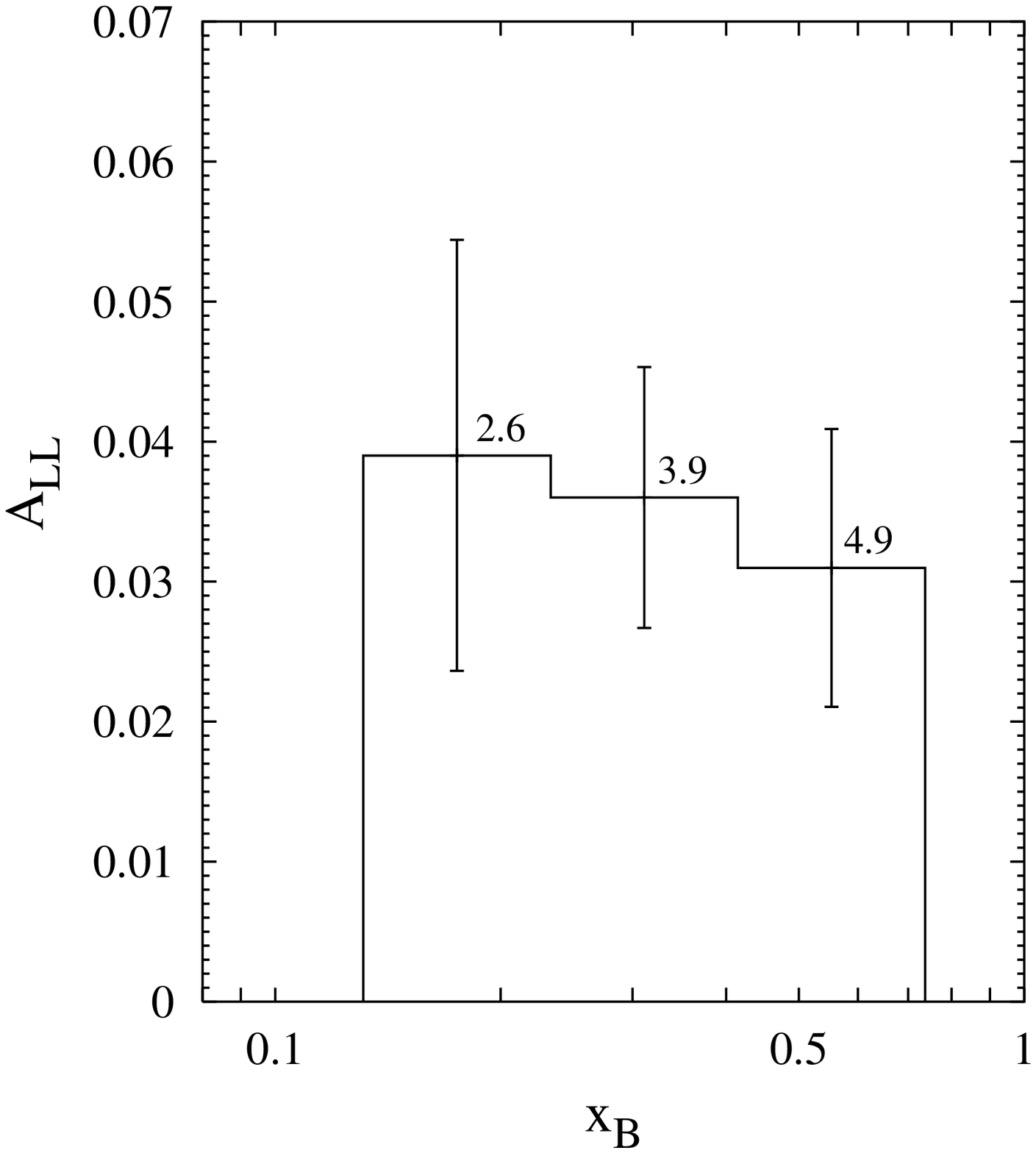,width=12.0cm,height=11.0cm}\\
\end{center}   
\begin{center} 
\parbox{14.0cm}
{{\footnotesize
Fig. 5: Asymmetry in the inelastic channel in bins of $x_B$ at HERMES. We
have used Badelek {\it et al.}  \cite{bad} parametrization of $g_1$. The
constraints imposed are as in Table I (except $\hat s > Q^2$), together with 
$\hat S - \hat s > 2 \,\,\mathrm{GeV}^2$. The average $Q^2$ 
(in $\mathrm{GeV}^2$) of  each bin is also shown.}}  
\end{center}
\newpage      
\begin{center}
\parbox{8cm}{\epsfig{figure=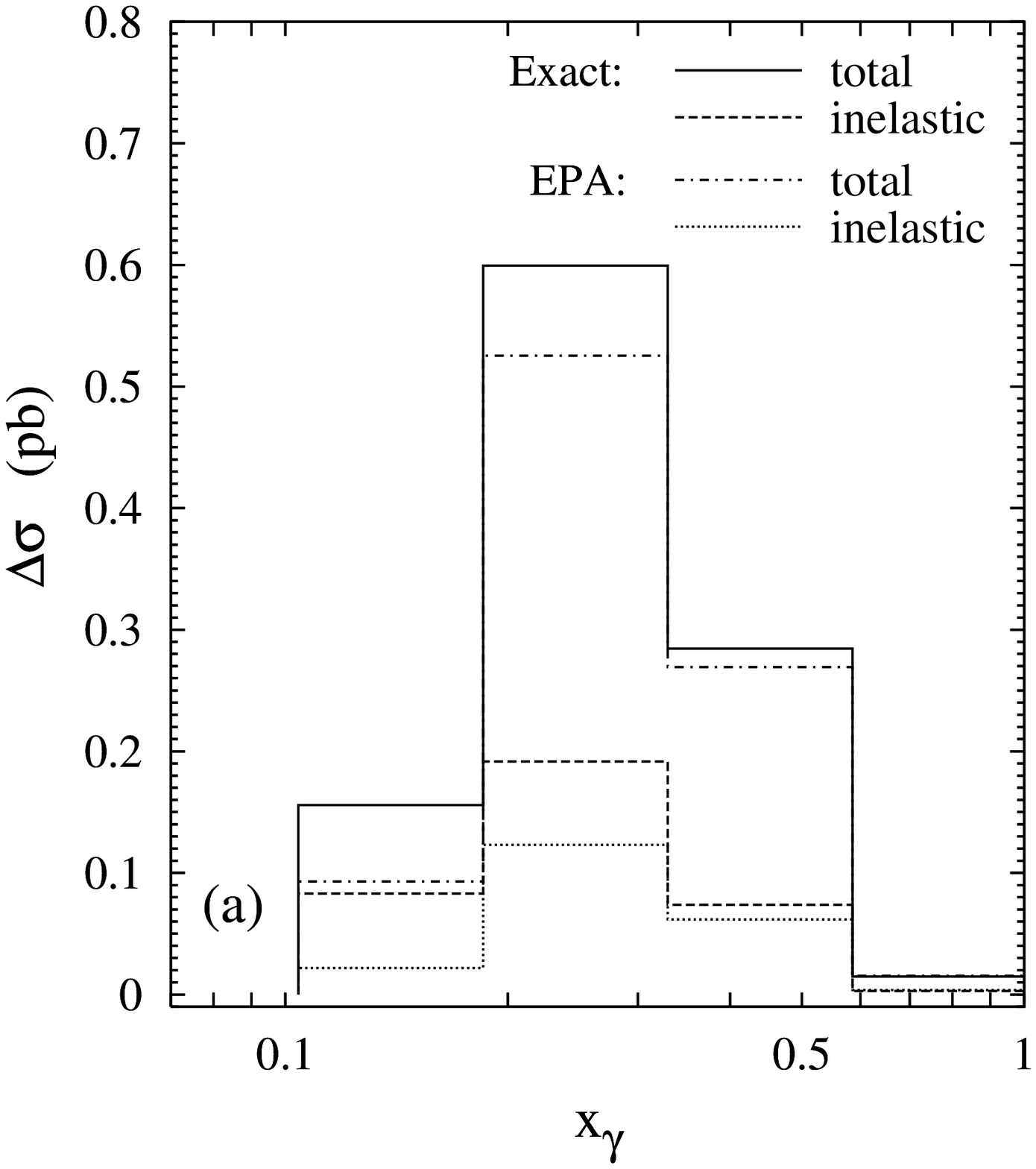,width=8.5 cm,height=7.5 cm}}\
\
\parbox{8cm}{\epsfig{figure=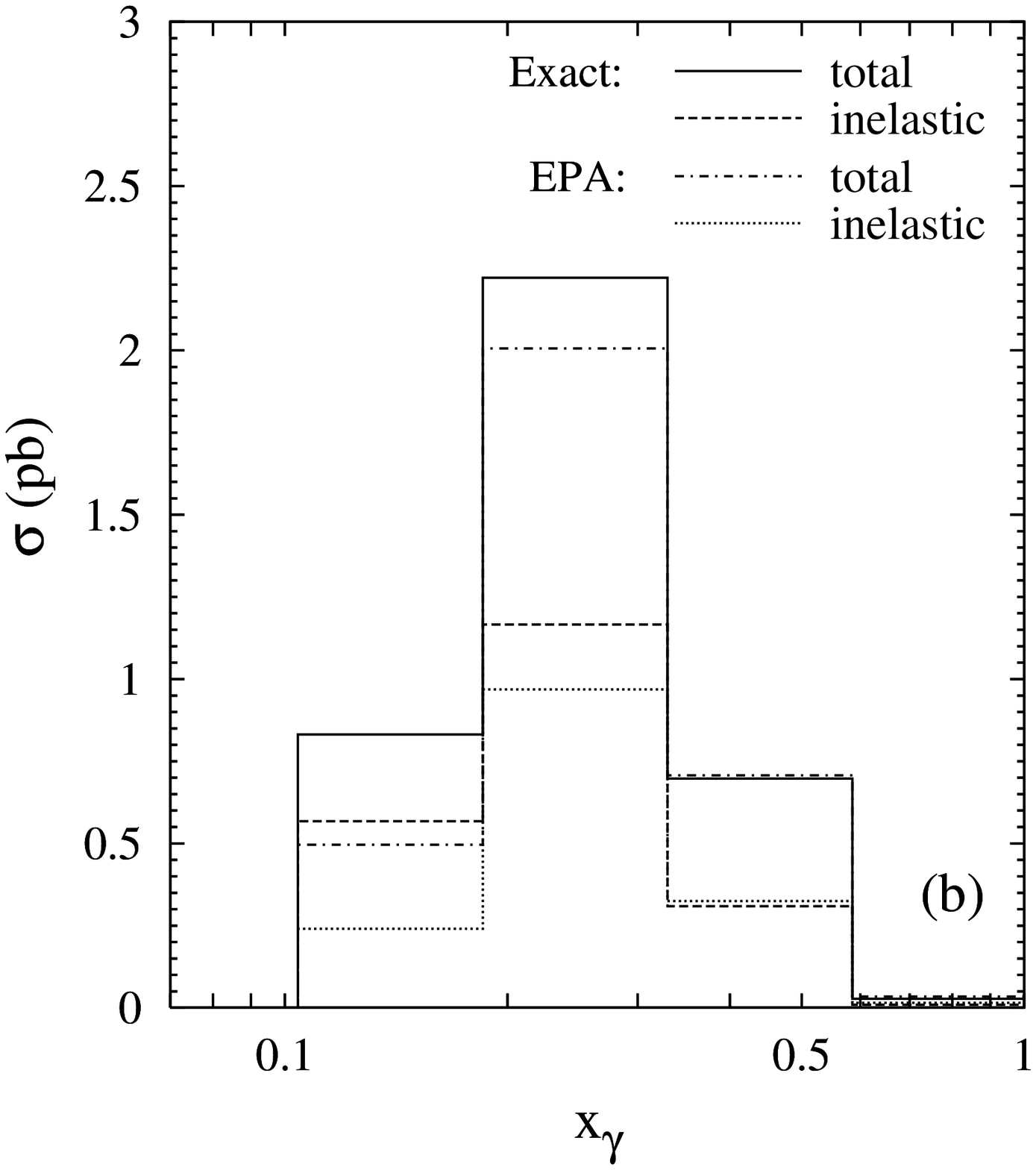,width=8.5 cm,height=7.5 cm}}\
\
\end{center}   
\vspace{0.2cm}

\begin{center}
\parbox{8cm}{\epsfig{figure=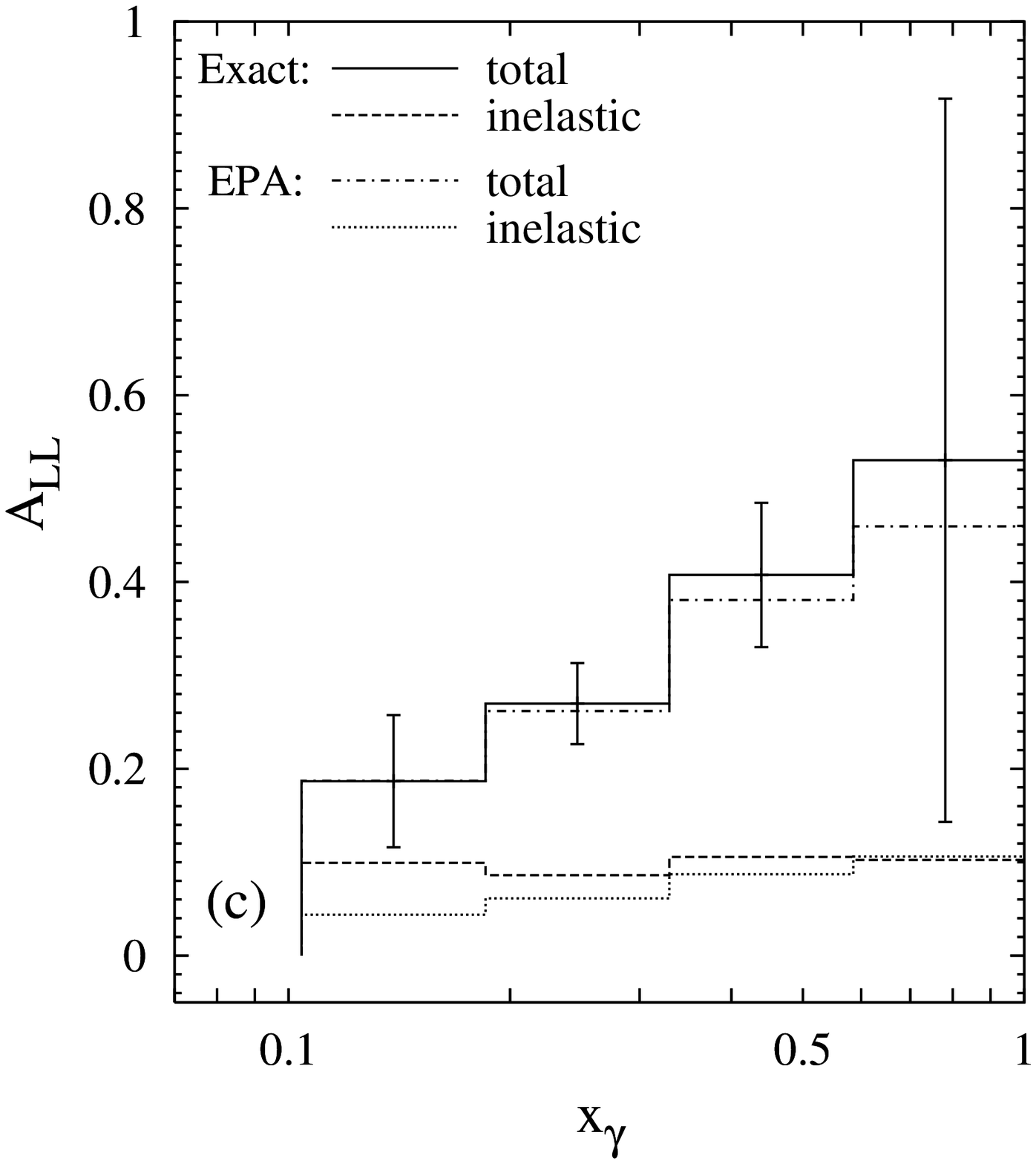,width=8.5 cm,height=7.5 cm}}\
\
\parbox{8cm}{\epsfig{figure=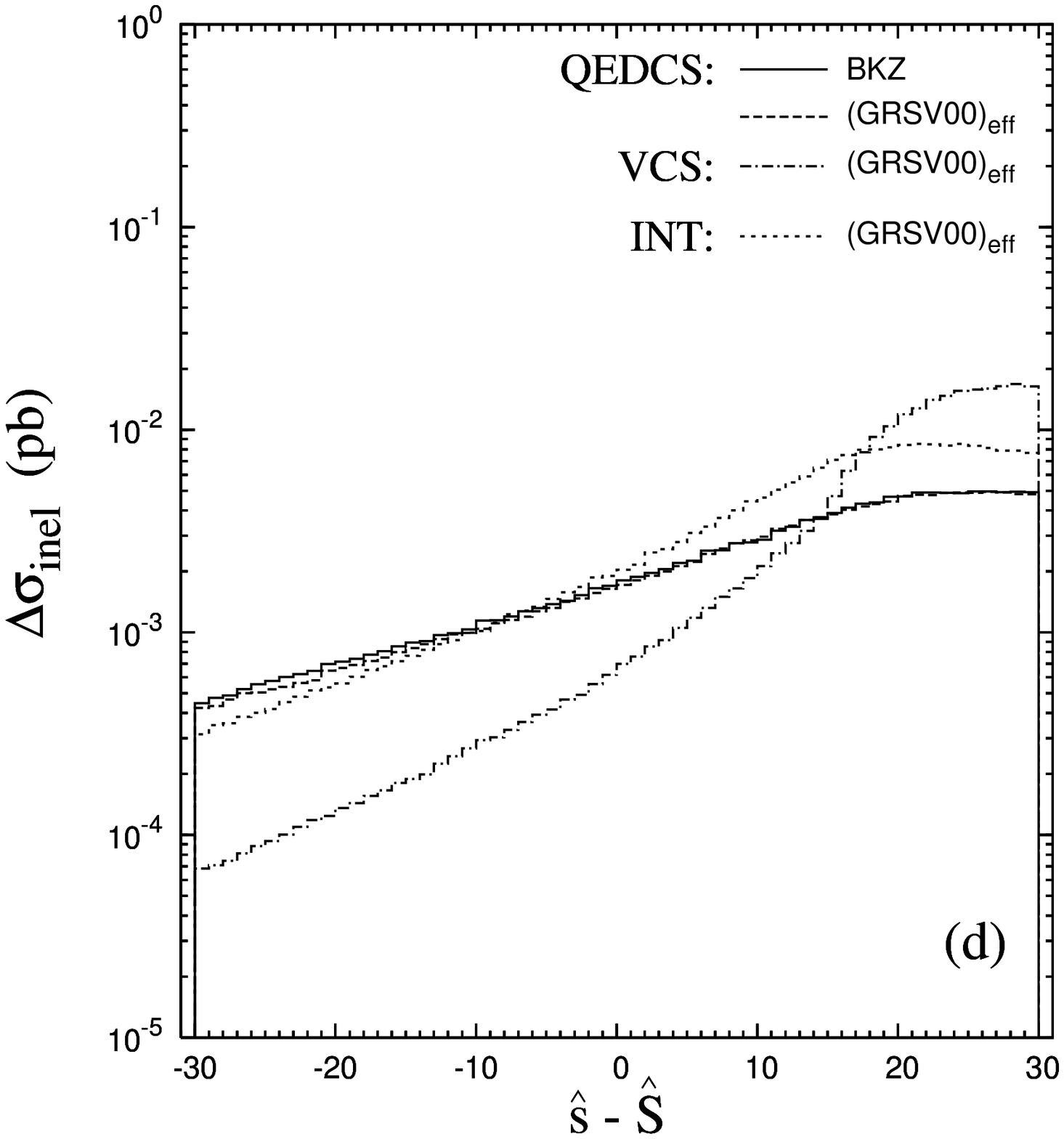,width=8.5 cm,height=7.5 cm}}\
\
\end{center}   
\vspace{0.2cm}
\begin{center}
\parbox{14.0cm}
{{\footnotesize
Fig. 6:  (a), (b), (c) and (d) are the same as in Fig. 4 but for COMPASS. The
constraints imposed are given in  Table I.}} 
\end{center}
\newpage
\vspace*{4cm} 
\begin{center}
\epsfig{figure= 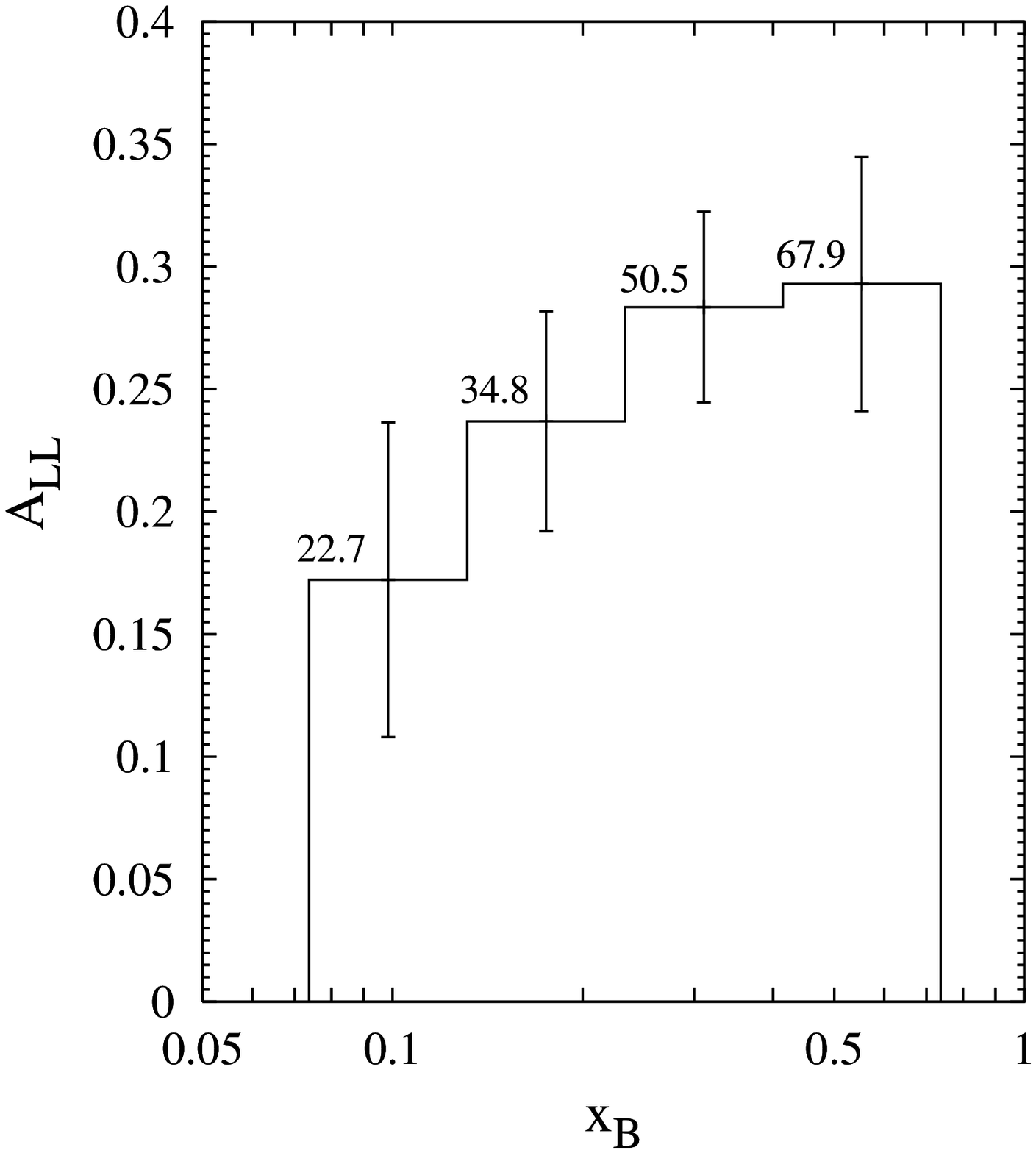,width=12.0cm,height=11.0cm}\\
\end{center}   
\begin{center} 
\parbox{14.0cm}
{{\footnotesize
Fig. 7: Asymmetry in the inelastic channel in bins of $x_B$ at COMPASS.We
have used Badelek {\it et al.} \cite{bad} parametrization of $g_1$. The constraints imposed are as in  Table I (except $\hat s > Q^2$), together with $\hat S - \hat s > 2 \,\, \mathrm{GeV}^2$. The average $Q^2$ (in $\mathrm{GeV}^2$) of  each
 bin is also shown.}}  
\end{center}

\newpage      
\begin{center}
\parbox{8cm}{\epsfig{figure=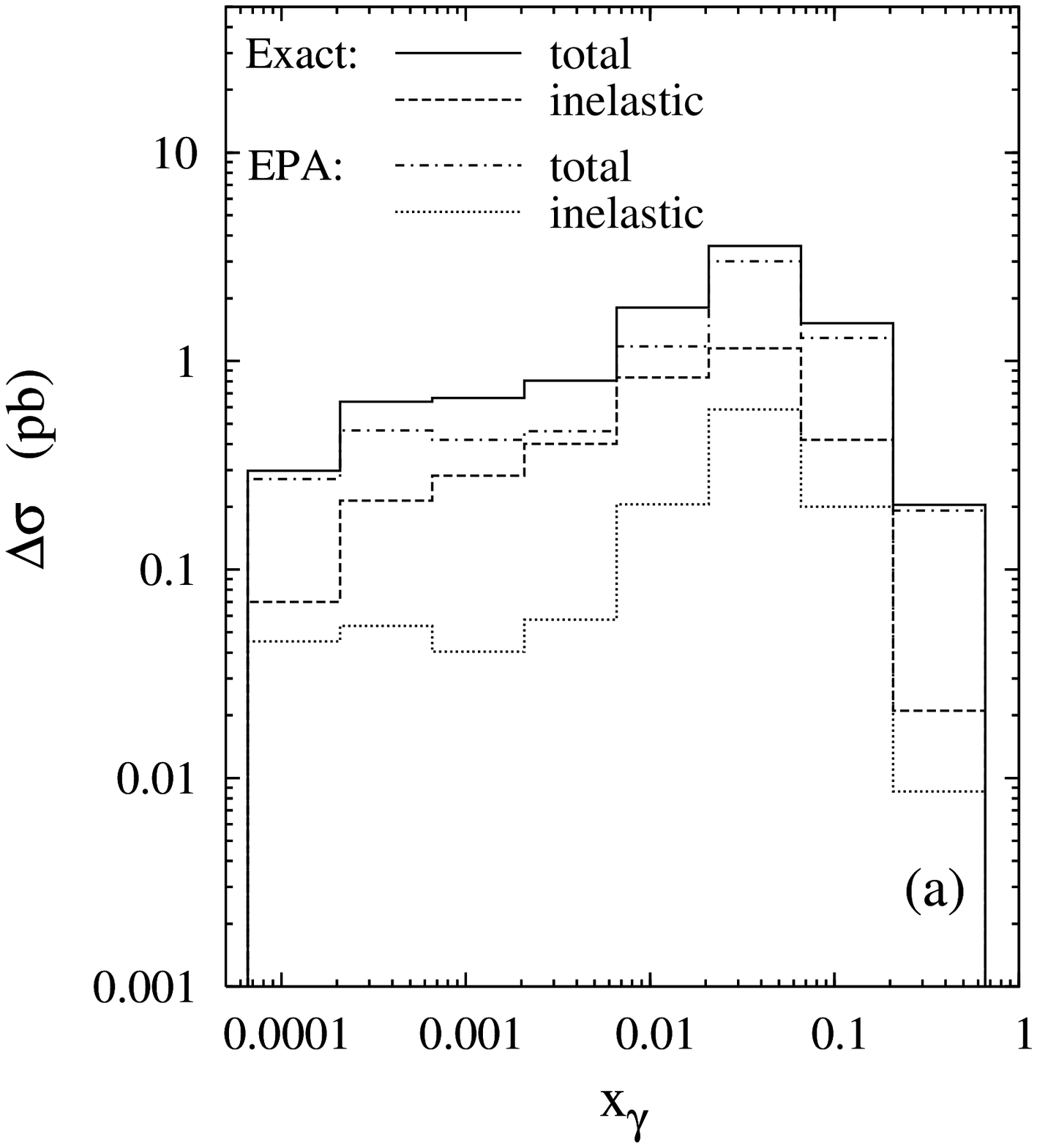,width=8.5 cm,height=7.5 cm}}\
\
\parbox{8cm}{\epsfig{figure=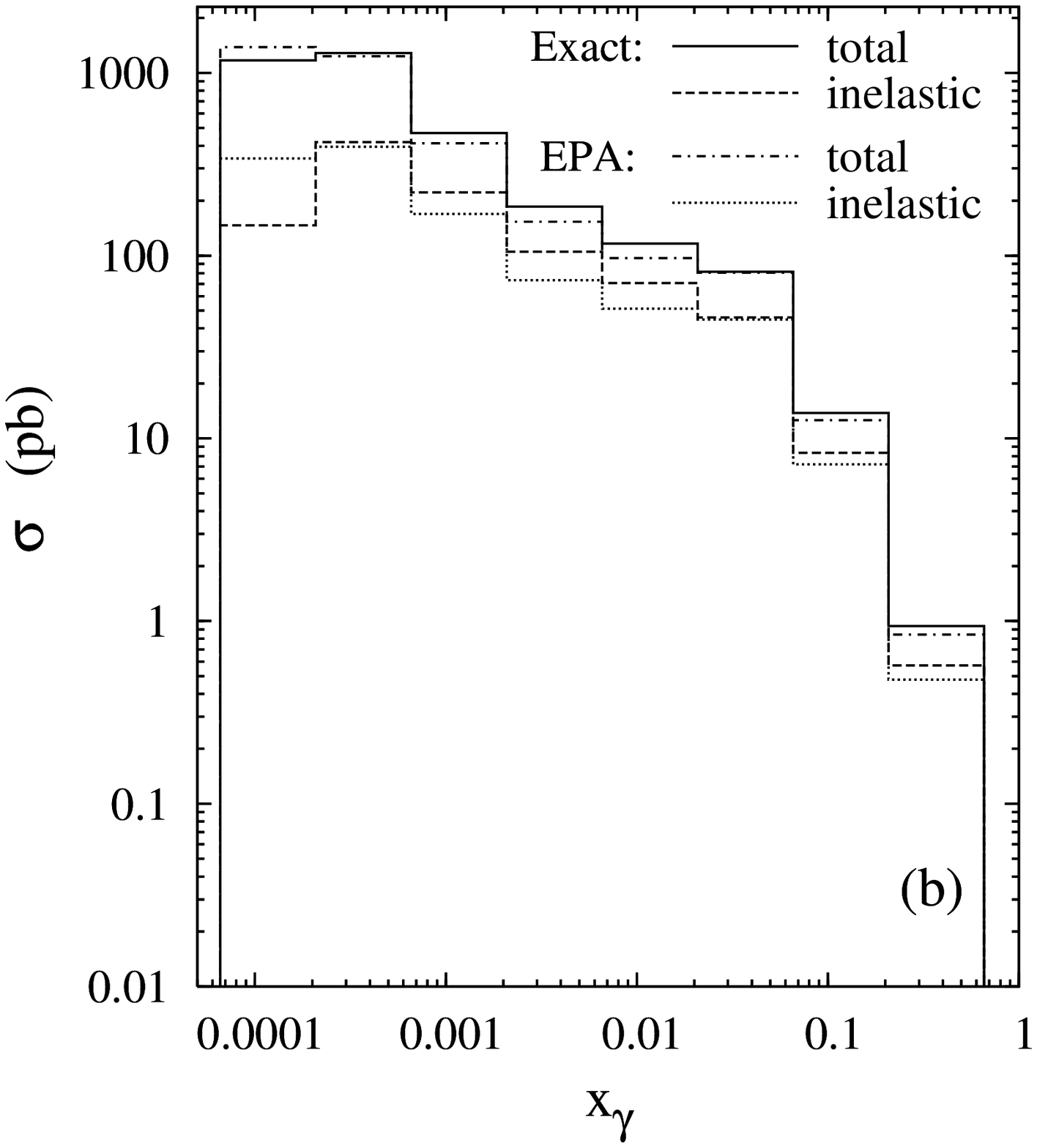,width=8.5 cm,height=7.5 cm}}\
\
\end{center}   
\vspace{0.2cm}

\begin{center}
\parbox{8cm}{\epsfig{figure=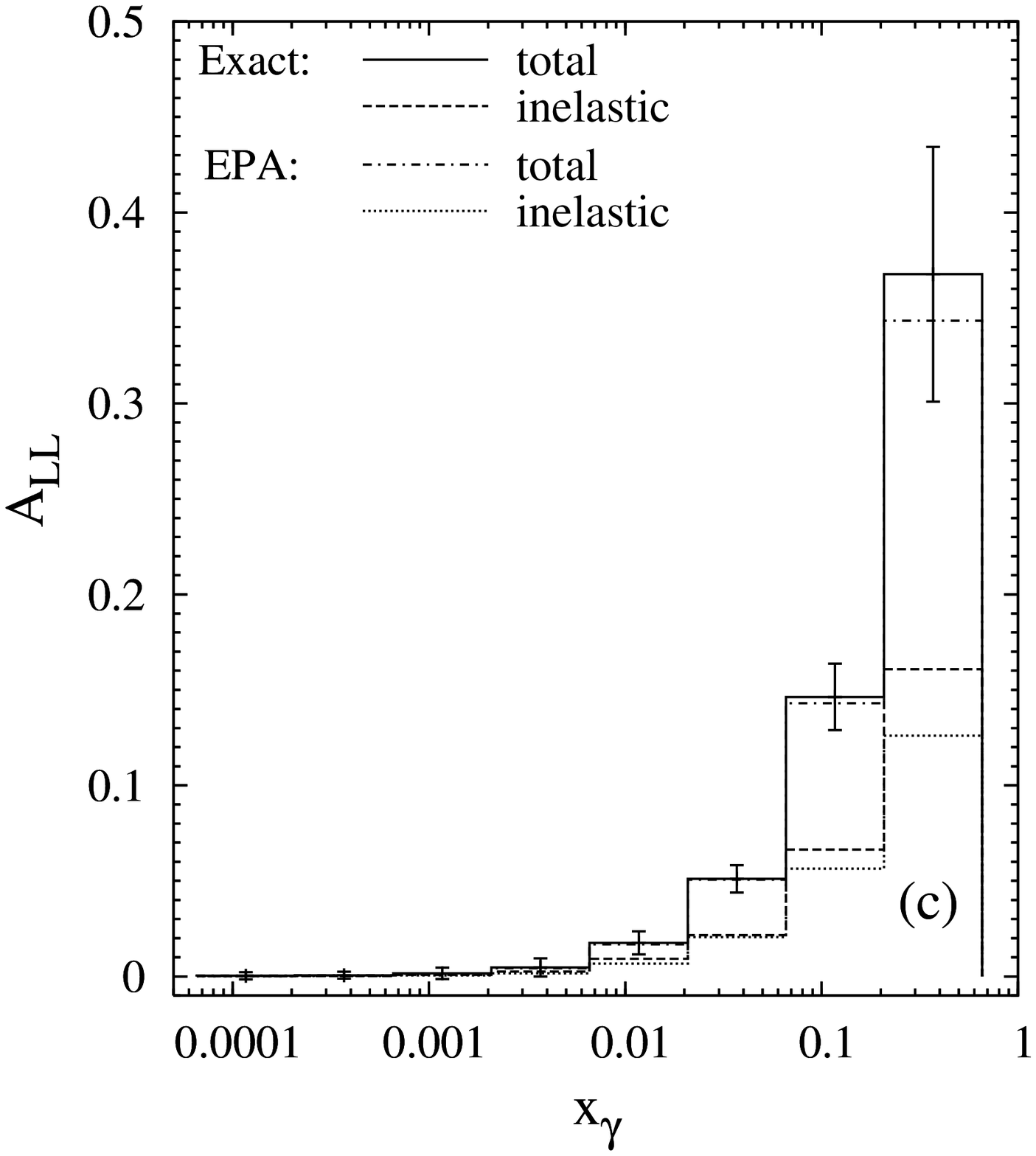,width=8.5 cm,height=7.5 cm}}\
\
\parbox{8cm}{\epsfig{figure=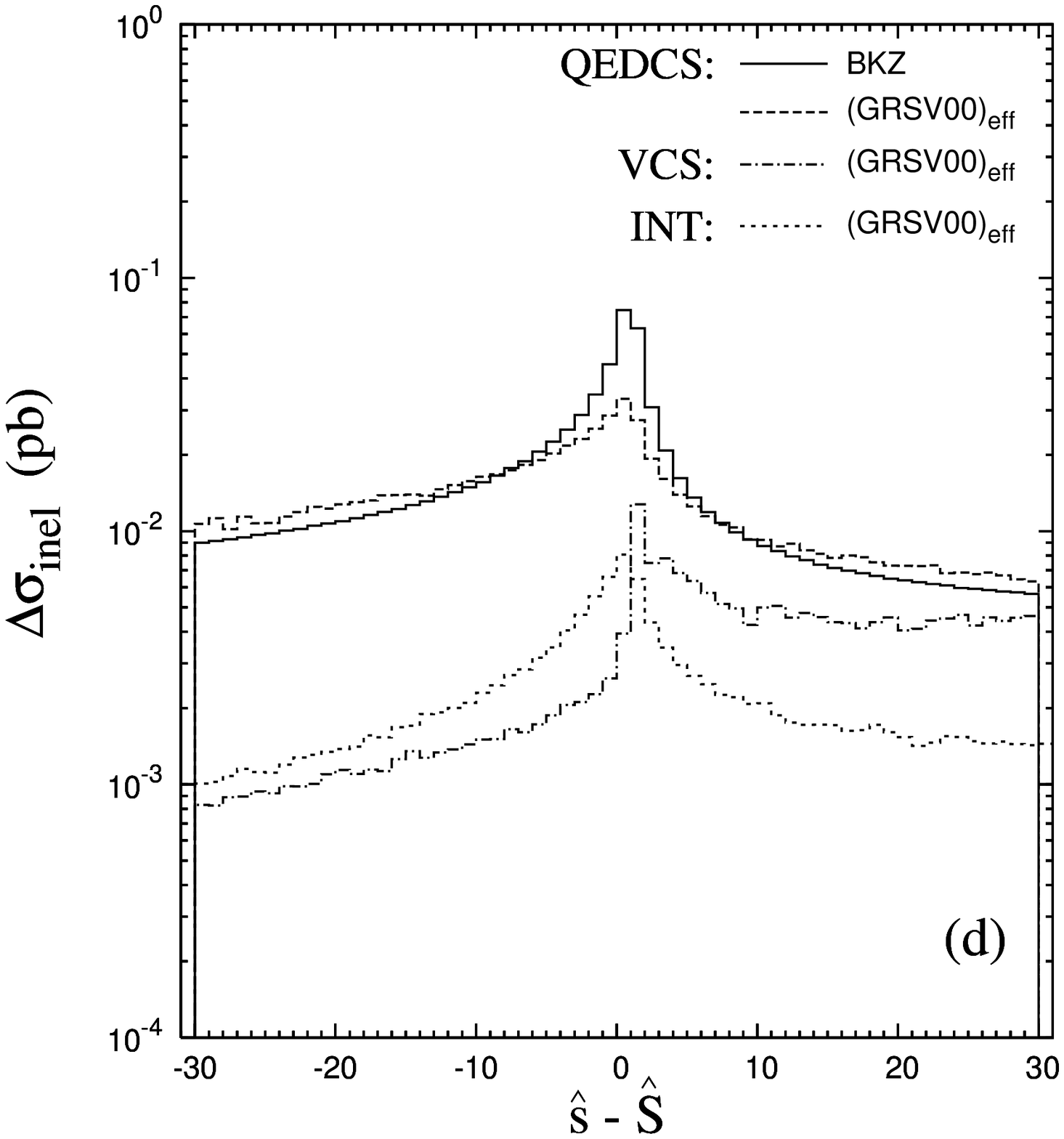,width=8.5 cm,height=7.5 cm}}\
\
\end{center}   
\vspace{0.2cm}
\begin{center}
\parbox{14.0cm}
{{\footnotesize
Fig. 8:  (a), (b), (c) and (d) are the same as in Fig. 4 but for eRHIC. The
constraints imposed are given in Table I.}} 
\end{center}
\newpage
\vspace*{4cm} 
\begin{center}
\epsfig{figure= 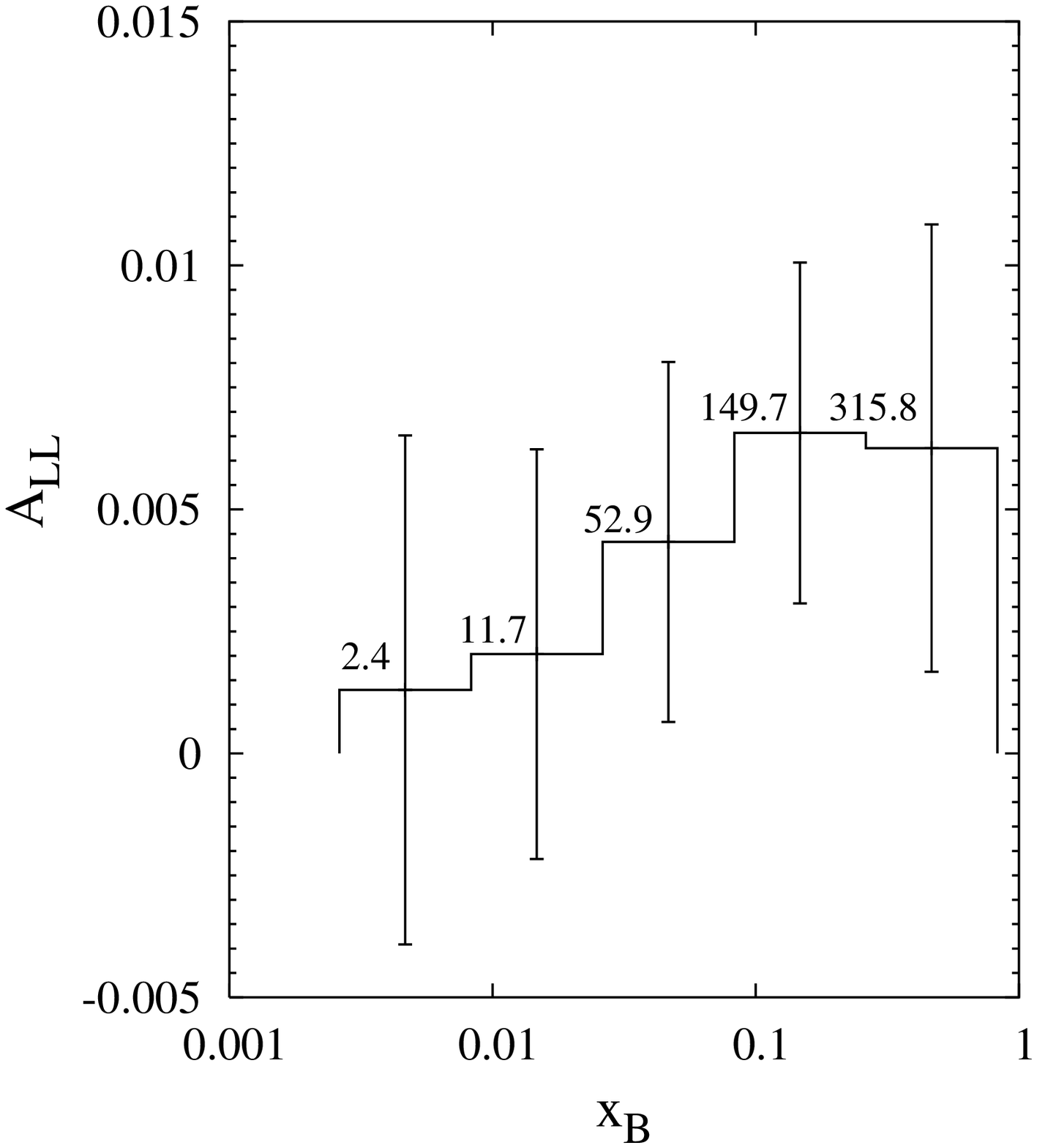,width=7.0cm,height=7.0cm}\\
\end{center}   
\begin{center} 
\parbox{14.0cm}
{{\footnotesize
Fig. 9: Asymmetry in the inelastic channel in bins of $x_B$ at eRHIC. We
have used Badelek {\it et al.} \cite{bad} parametrization of $g_1$. The
constraints imposed are as in Table I (except $\hat s > Q^2$), together with 
$\hat S > \hat s$. The average $Q^2$ (in $\mathrm{GeV}^2$) of  each
 bin is also shown.}}  
\end{center}
\vspace*{.2cm} 
\begin{center}
\epsfig{figure= 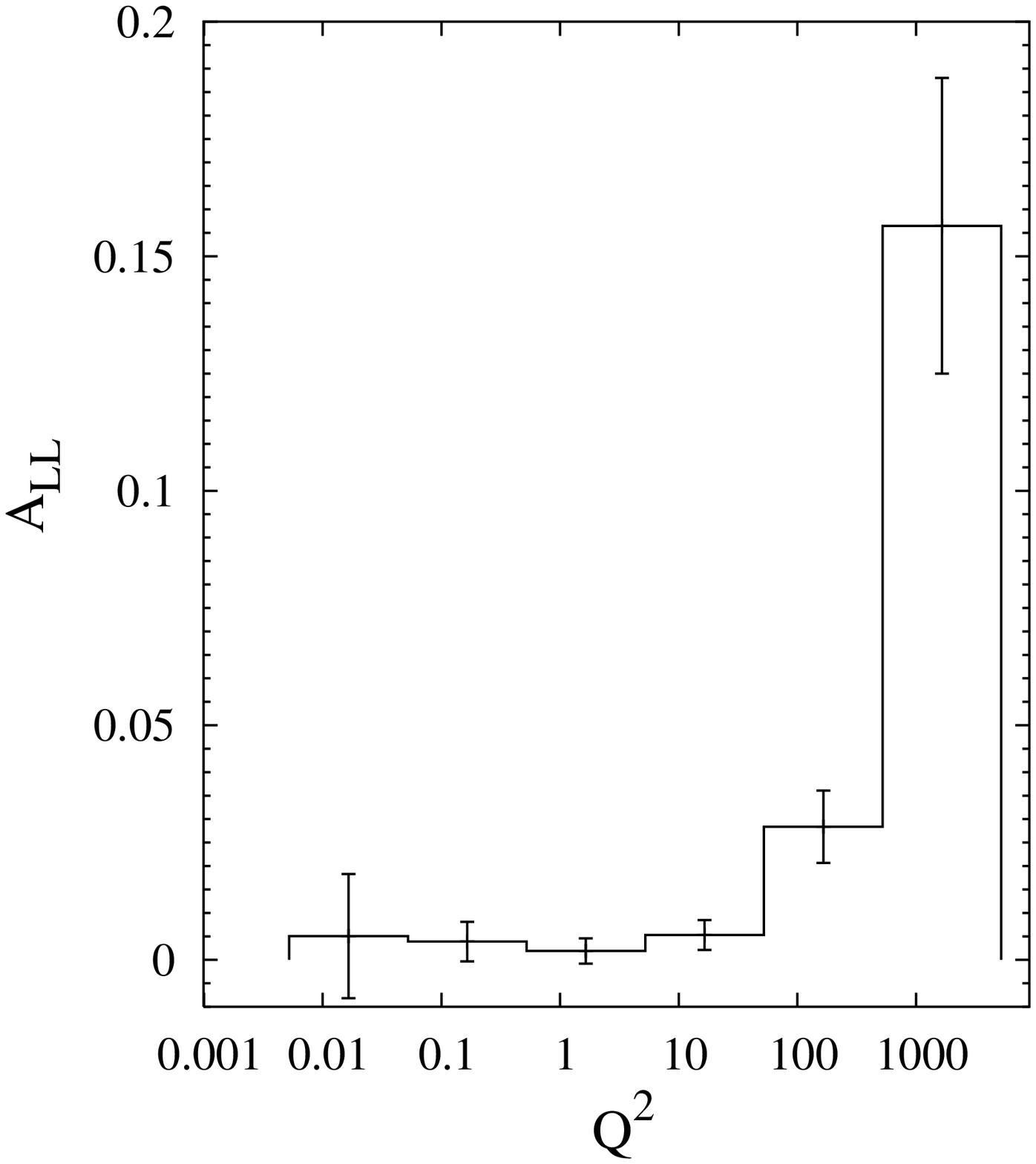,width=7.0cm,height=7.0cm}\\
\end{center}   
\begin{center} 
\parbox{14.0cm}
{{\footnotesize
Fig. 10: Asymmetry  in bins of $Q^2$ ($\mathrm{GeV}^2$) at eRHIC. We
have used Badelek {\it et al.} \cite{bad} parametrization of $g_1$. The
constraints imposed are the same as in Fig. 9.}}  
\end{center}
\newpage
\begin{table}[c]
\begin{center}
\vspace*{8cm}
\hspace*{-0.5cm}
\begin{tabular}{|c|c|c|}
\hline 
 HERMES  & COMPASS  & eRHIC   \\
\hline\hline 
& & $~~E_{p} = 250\, \mathrm{GeV}~~$ \\
$~~E_e = 27.5 \, \mathrm{GeV}~~ $ & $~~E_{\mu} =  160 \,\mathrm{GeV}~~ $ & $E_{e} = 10\, \mathrm{GeV}~~$ \\  $~~~ 0.04 < \theta_e, \, \theta_{\gamma} <  0.2~~~$ &$~~~ 0.04 < \theta_{\mu}, \, \theta_{\gamma} <  0.18~~~$    &
$~ 0.06 < \theta_e, \, \theta_{\gamma} < \pi - 0.06~$ \\
$~~E_e', \, E_{\gamma}' > 4 \, \mathrm{GeV}~~ $ & $~~E_{\mu}', \, E_{\gamma}' > 4 \,\mathrm{GeV}~~ $ &$~~E_e', \, E_{\gamma}' > 4 \, \mathrm{GeV}~~ $  \\
 $ \hat s > 1\,\mathrm{GeV}^2  $  & $ \hat s > 1\,\mathrm{GeV}^2  $ & $ \hat s > 1\,\mathrm{GeV}^2 $  \\
 $ \hat s > Q^2 $  & $ \hat s > Q^2 $ & $\hat s > Q^2 $  \\

\hline

\end{tabular}
\end{center}
\caption{Energies, angular acceptance and kinematical cuts for various experiments.}
\label{tableone}

\end{table}


\begin{thebibliography}{99}
\bibitem{blu} J. Bl\"umlein, G. Levman, H. Spiesberger, J. Phys. {\bf G 19},
1695 (1993). 
\bibitem{ruju} A. De Rujula, W. Vogelsang, Phys. Lett. {\bf B 451} 437
(1999). 
\bibitem{kessler} A. Courau and P. Kessler, Phys. Rev. {\bf D 46}, 117,
(1992).
\bibitem{lend} V. Lendermann, H. C. Schultz-Coulon, D. Wegener, Eur. Phys.
J. {\bf C 31} 343 (2003).
\bibitem{pap1} A. Mukherjee, C. Pisano, Eur. Phys. J. {\bf C 30}, 477
(2003).  
\bibitem{pp2} A. Mukherjee, C. Pisano, hep-ph/0402046, to appear in Eur.
Phys. J. C.
\bibitem{gpr1} M. Gl\"uck, C. Pisano, E. Reya, Phys. Lett. {\bf B 540}, 75,
(2002).
\bibitem{gpr2} M. Gl\"uck, C. Pisano, E. Reya, I. Schienbein,
Eur. Phys. J. {\bf C 27}, 427 (2003).
\bibitem{slac} K. Abe {\it et. al.}, Phys. Rev. {\bf D 58}, 112003 (1998).
\bibitem{her1} A. Airapetian {\it et. al.}, Phys. Lett. {\bf B 494}, 1
(2000).
\bibitem{her2} A. Airapetian {\it et. al.}, Phys. Rev. Lett. {\bf B 90},
092002, (2003).
\bibitem{clas1} M. Amarian {\it et. al.}, Phys. Rev. Lett. {\bf 89}, 242301
(2002).
\bibitem{clas2} J. Yun {\it et. al.}, Phys. Rev. {\bf C 67}, 055204, (2003).
\bibitem{clas3} R. Fatemi {\it et. al}, Phys. Rev. Lett. {\bf 91}, 222002,
(2003). 
\bibitem{badel} B. Badelek, Acta Phys. Pol. {\bf B  34}, 2943 (2003)
\bibitem{reya} B. Lampe and E. Reya, Phys. Rept. {\bf 332}, 1 (2000). 
\bibitem{bass} S. D. Bass and  A. De Roeck, Nucl. Phys. B  (Proc. Suppl.) {\bf 105}, 1 (2002) 
\bibitem{kniehl} B. Kniehl, Phys. Lett. {\bf B 254}, 267, (1991).  
\bibitem{flo} D. de Florian, S. Frixione, Phys. Lett. {\bf B 457}, 236,
(1999).
\bibitem{ji} X. Ji, Phys. Rev. {\bf D 55}, 7114 (1997).
\bibitem{gpd} For reviews on generalized parton distributions, see M. Diehl,
Phys. Rept, {\bf 388}, 41 (2003); X. Ji, J. Phys. {\bf G 24}, 1181 (1998);
A. V. Radyushkin,
hep-ph/0101225, published in "At the Frontier of Particle Physics/Handbook
of QCD", ed. M. Shifman (World Scientific, Singapore, 2001); K. Goeke, M.
V. Polyakov, M. Vanderhaeghen, Prog. Part. Nucl. Phys. {\bf 47}, 401 (2001). 
\bibitem{thesis} V. Lenderman, Ph. D. thesis, Univ. Dortmund,
H1 collaboration, DESY-THESIS-2002-004, (2002).
\bibitem{allm} H. Abramowicz and A. Levy, hep-ph/9712415; corrected according
to a private communication by the authors.
\bibitem{bad} B. Badelek, J. Kwiecinski and B. Ziaja, Eur. Phys. J. {\bf C
26}, 45 (2002).
\bibitem{grsv} M. Gl\"uck, E. Reya, M. Stratmann and W. Vogelsang, Phys.
Rev. {\bf D 63}, 094005 (2001).
\end{thebibliography}
\end{document}